\def\ep{E_{\rm peak}}

\def\eps{E^{\rm src}_{\rm peak}}

\documentclass[12pt,preprint]{aastex}

\slugcomment{Not to appear in Nonlearned J., 45.}

\shorttitle{Gamma-Ray;burst}
\shortauthors{Sakamoto et al.}

\begin{document}


\title{Evidence of 
Exponential Decay Emission \\in the {\it \bfseries Swift}  Gamma-ray Bursts}


\author{T. Sakamoto\altaffilmark{1,2}, 
J. E. Hill\altaffilmark{1,3,4},
R. Yamazaki\altaffilmark{5},
L. Angelini\altaffilmark{1},
H. A. Krimm\altaffilmark{1,3,4},
G. Sato\altaffilmark{1,6},\\
S. Swindell\altaffilmark{7}, 
K. Takami\altaffilmark{5}, 
J. P. Osborne\altaffilmark{8}}

\altaffiltext{1}{NASA Goddard Space Flight Center, Greenbelt, MD 20771}
\altaffiltext{2}{Oak Ridge Associated Universities, P.O. Box 117, 
 Oak Ridge, Tennessee 37831-0117}
 \altaffiltext{3}{CRESST NASA Goddard Space Flight Center, Greenbelt, MD 20771}
\altaffiltext{4}{Universities Space Research Association, 10211 Wincopin 
	Circle, Suite 500, Columbia, MD 21044-3432} 
\altaffiltext{5}{Department of Physics, Hiroshima University, Higashi-Hiroshima, 
	Hiroshima, 739-8526, Japan}
\altaffiltext{6}{Institute of Space and Astronautical Science, 
JAXA, Kanagawa 229-8510, Japan}
\altaffiltext{7}{Department of Physics, North Carolina Agricultural
and Technical State University, 1601 East Market Street,Greensboro, North Carolina 27411}
\altaffiltext{8}{Department of Physics and Astronomy, University of
Leicester, LE1, 7RH, UK}



\begin{abstract}

We present a systematic study of the steep decay emission 
from gamma-ray bursts (GRBs) observed by the {\it Swift} X-Ray 
Telescope (XRT).  In contrast to the analysis described in recent 
literature, we produce composite Burst Alert Telescope (BAT) and 
XRT light curves by extrapolating the XRT data (2--10 keV) into 
the BAT energy range (15--25 keV) rather than extrapolating the 
BAT data into the XRT energy band (0.3--10 keV).  Based on the fits 
to the composite light curves, we have confirmed the existence of 
an exponential decay component which smoothly connects the BAT 
prompt data to the XRT steep decay for several GRBs.  We also 
find that the XRT steep decay for some of the bursts can be well 
fit by a combination of a power-law with an exponential decay model.  
We discuss this exponential component within the frame work of both 
the internal and the external shock model.  

\end{abstract}





\section{Introduction}

The transition between the gamma-ray burst (GRB) prompt emission and 
afterglow emission has generated great interest within the scientific 
community.  It is generally accepted that the GRB prompt
emission is due to internal shocks originating from the collision 
of faster and slower moving shells, whereas, the afterglow is believed 
to originate from an external shock resulting from 
the relativistic fireball colliding with a circum-burst medium 
\citep{rm1994,sp1997,mr1997}.  During the GRB episode there should be a
transition from one phase to the other, however, it is still 
not well understood as to when this transition occurs.  
The Burst and Transient Source Experiment (BATSE) observation 
of GRB 980923 showed long lasting tail emission, $\sim$400 s, which 
is best described by a power-law temporal decay \citep{giblin1999}.  
Based on the spectral and temporal characteristics of this burst, 
\citet{giblin1999} concluded that the tail emission was a part of 
the afterglow emission, thus, the external shock could be generated 
during the prompt $\gamma$-ray phase.  \citet{giblin2002} performed 
a systematic study of the prompt tail emission using 40 GRBs observed 
by BATSE.  They found that the temporal decays are best described by a 
power-law with a decay index of $-2$ rather than an exponential.  
There are several other analysis of BATSE data which reach the same 
conclusion \citep[e.g.,][]{rs2002}.  
According to the BeppoSAX observations, the late time afterglow smoothly 
connects with the prompt emission if the onset time of the light curve 
is defined as the start time of the last pulse observed in the Wide 
Field Camera (2--30 keV) \citep[e.g.,][]{pian2001,piro2005}.  
These observations support the idea that the late X-ray spike represents 
the onset of the afterglow.  However, the delay of a few hours to a 
few days before the narrow field X-ray instrument is pointed to the 
GRB position, weakens the discussion concerning the transition from 
the prompt emission to the afterglow.  

As a result of the revolutionary features of {\it Swift} 
\citep{gehrels2004}, our understanding of the X-ray properties of 
GRBs has been improved dramatically.  With the combination of the 
accurate on-board calculation of the GRB position by the Burst Alert Telescope 
\citep[BAT:15-150 keV;][]{barthelmy2005} and the fast slewing capability 
of the spacecraft, {\it Swift} can begin a highly sensitive X-ray 
observation with the X-Ray Telescope \citep[XRT:0.2-10 keV;][]{burrows2005} 
within a few tens of seconds to a few hundred seconds after the 
burst trigger.  According to the XRT observations, the X-ray properties 
of the GRB emission have very complex features
\citep{nousek2006,zhang2006a}.  One of the most unexpected discoveries 
by the XRT is the existence of the steep decay component during the 
initial phase of the X-ray light curve.  The origin of this steep decay 
component is generally considered to be a result of the delayed 
prompt emission from different viewing latitudes of the jet, the so 
called ``curvature effect'' \citep[e.g.][]{fenimore1996,kumar2000,zhang2006a}.  
\citet{tagliaferri2005} and \citet{barthelmy2006} 
investigated the steep decay component with the composite BAT and XRT 
light curves for several GRBs.  To generate the composite light curve, 
both papers performed an extrapolation of the BAT mask-weighted 
(background subtracted) light curve into the XRT 0.2--10 keV energy band 
using a best-fit power-law photon index.  The authors found that GRB 
050126 and GRB 050219A do not show continuous emission from the BAT to 
the XRT light curve, however, GRB 050315 and GRB 050319 do display 
a smooth continuation from the BAT to the XRT light curve. 
\citet{obrien2006} performed a systematic study of the early X-ray emission 
using a sample of 40 {\it Swift} GRBs.  They constructed a composite BAT 
and XRT light curve in the 0.3--10 keV band.  In order to extrapolate 
the BAT data points, the BAT mask-weighted count rate was converted to flux 
in the 0.3--10 keV band using the mean of the best-fit photon indices 
obtained from a simple power-law fit to both the BAT and the XRT spectrum.  
The authors found that a fit of the 40 superimposed GRB light curves 
in the 0.3-10 keV band could be described by an exponential decay 
followed by a power-law decay.  \citet{willingale2007} investigated the 
X-ray emission including the late time light curve data.  
They found that the X-ray light curve data can be modeled with the 
superposition of an early (``prompt'') and a late time (``afterglow'')
component.  

\begin{figure}
\centerline{
\includegraphics[width=8.5cm]{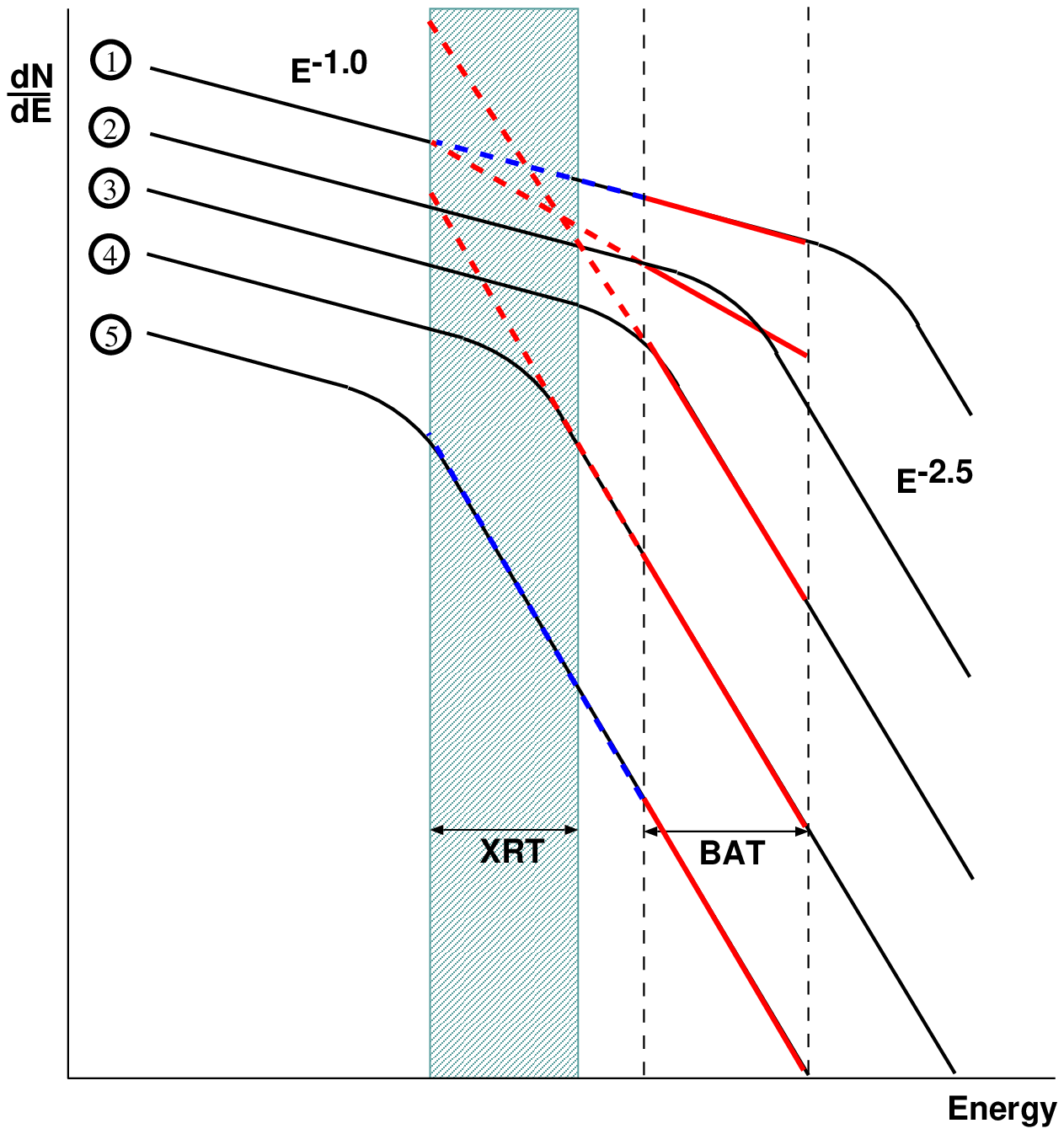}
\vspace{0.5cm}
\includegraphics[width=8.5cm]{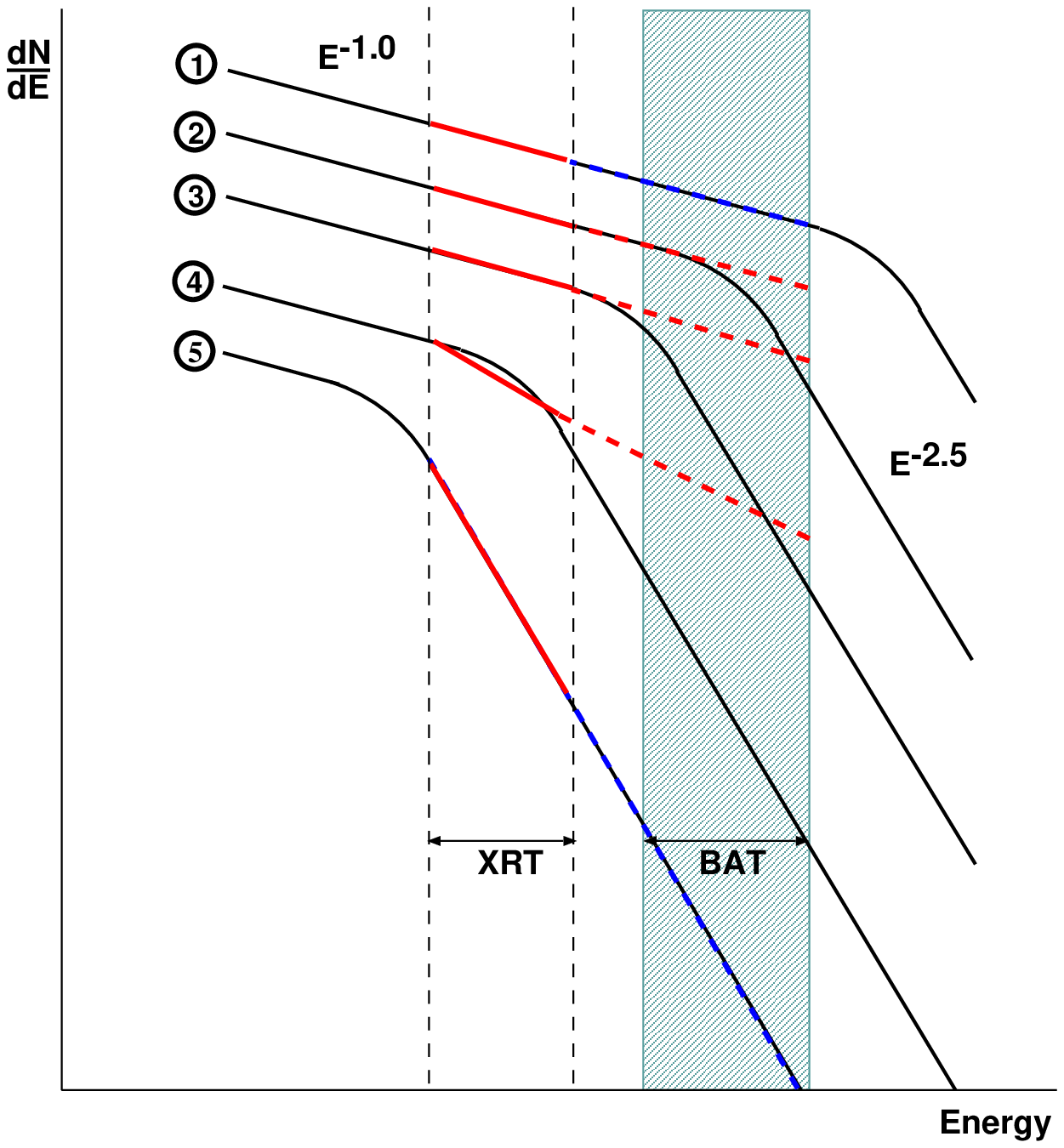}}
\caption{Schematic figures of the observed spectra with different 
 $\ep$ energies (case 1: $\ep > 150$ keV, case 2: 15 keV $< \ep <$ 150 keV, 
case 3: 10 keV $< \ep <$ 15 keV, case 4: 0.3 keV $< \ep <$ 10 keV, 
and case 5: $\ep <$ 0.3 keV) demonstrating issues with the different 
 extrapolations (left: extrapolating the BAT data down to the XRT energy
 band (BAT-to-XRT extrapolation); 
right: extrapolating the XRT data up to the BAT energy range (XRT-to-BAT 
 extrapolation)).  
The light blue hatched regions show the extrapolated energy band.  
The extrapolated spectra shown with red dotted lines indicate issue with 
the extrapolation.  There should be no issue with the extrapolation 
of the spectra shown in case 1 and 5 (blue dotted lines).  See text for 
details  (section 1).}
\label{fig:prob_extrapolation_bat_xrt}
\end{figure}

In this paper, we describe the analysis 
of the tail emission using an alternative approach. 
We generate the composite BAT and XRT light curve in the 15--25 keV BAT 
energy band by extrapolating the 2--10 keV XRT count rate into the 
15-25 keV band.  Either approach, extrapolating the BAT count rate down 
to the XRT band (hereafter BAT-to-XRT extrapolation) or extrapolating 
the XRT count rate up to the BAT band (hereafter XRT-to-BAT extrapolation), 
can encounter similar systematic problems.  Most of the GRB prompt 
emission spectra is well fit by the Band function \citep{band1993} with 
a low-energy and a high-energy photon index of $\sim 1$ and $\sim 2.3$, 
respectively \citep[e.g.][]{sakamoto2005,kaneko2006}.  However, it is a 
well known characteristic that the peak energy of the spectrum, $\ep$, 
not only evolves during the burst \citep[e.g.,][]{lloyd2002,frontera1999} 
but also changes from burst-to-burst \citep[e.g.,][]{sakamoto2005}.  
Here, we discuss the issues with regard to both types of extrapolation
by considering five cases depending on the value of $\ep$ 
(shown in figure \ref{fig:prob_extrapolation_bat_xrt} left is the BAT-to-XRT 
extrapolation and right is the XRT-to-BAT extrapolation).  If we 
consider the energy bands of the BAT (15-150 keV) and the XRT (0.3-10 keV), 
the five cases 
are defined by: 1) $\ep > 150$ keV, 2) 15 keV $< \ep <$ 150 keV, 
3) 10 keV $< \ep <$ 15 keV, 4) 0.3 keV $< \ep <$ 10 keV, and 5) $\ep <$ 0.3 keV.  
For cases 1 and 5, both the BAT-to-XRT and the XRT-to-BAT extrapolations 
should provide the correct flux because the photon index of the
extrapolated energy band is the same as the observed energy band 
(``1'' and ``5'' of the left and the right panels of figure 
\ref{fig:prob_extrapolation_bat_xrt}).  
For cases 3 and 4, the BAT-to-XRT extrapolation would over estimate 
the flux in the XRT energy band, since the photon index in the observed 
energy band is steeper 
than that of the extrapolated energy band (``3'' and ``4'' of the 
left panel of figure \ref{fig:prob_extrapolation_bat_xrt}).  
Similarly, for case 2 and 3, the XRT-to-BAT extrapolation would 
over estimate the flux in the BAT energy band, because the photon 
index in the observed energy band is shallower than that of the 
extrapolated energy band (``2'' and ``3'' of the right panel of figure 
\ref{fig:prob_extrapolation_bat_xrt}).  
For case 2, using the BAT-to-XRT extrapolation (``2'' of the left 
panel of figure \ref{fig:prob_extrapolation_bat_xrt}), and for case 4, 
using the XRT-to-BAT  extrapolation (``4'' of the right panel of figure 
\ref{fig:prob_extrapolation_bat_xrt}), 
the flux would be overestimated in the XRT and the BAT energy band respectively, 
because the simple power-law fit is 
a tangential line to the curved spectrum as a result of the narrow 
energy bands of the BAT and the XRT (e.g. Sakamoto et al. in
preparation).   

To date, most of the BAT and XRT composite light curves in the 
literature have been produced from a BAT-to-XRT extrapolation.  
However, we believe that the XRT-to-BAT extrapolation described here  
may minimize the systematic errors, especially with respect to 
investigating GRB tail emission, for the following three reasons.  
First, we can account for the spectral evolution during the prompt 
emission if we extract the flux from the time-resolved spectral 
analysis of the BAT data.  As previously mentioned, $\ep$ shifts 
from hundreds of keV to a few keV during the prompt emission.  If in 
the analysis, one does not account for the spectral evolution 
during the prompt emission, which is the case for the majority of the 
published XRT and BAT composite light curves using the BAT-to-XRT extrapolation, 
the systematic error in the extrapolated flux in the X-ray range 
could be significant.  Second, since the BAT mask-weighted 
count rate does not correct for the energy dependence of each photon, 
the count rate of the source in the off-axis direction will be 
systematically smaller than the on-axis case.  This effect becomes 
an issue when the BAT-to-XRT extrapolation has been performed by 
converting the BAT mask-weighted count-rate into flux using a fixed 
photon index obtained from a simple power-law fit.  According to the 
BAT Crab observation, the count rate from the Crab is $\sim$15\% smaller 
in the 45 degree off-axis case.  This off-axis effect is correctly 
taken into account in the BAT energy response matrix, but not in 
the BAT mask-weighted count rates.  Therefore, unless one applies an
additional off-axis correction to the BAT mask-weighted count rates, a 
systematically smaller flux is obtained if the source is in the off-axis
direction, which is always the case for the BAT GRB data prior to the 
spacecraft slewing to the GRB position.  Third, according to the 
GRB synchrotron shock model \citep{sari1998b}, if the observed spectrum 
has a photon index steeper than $2$, and a power-law index of an electron 
distribution, $p$, in the range of 2 $\le$ $p$ $<$ 3 (where
N$(\gamma_{e}) \propto \gamma_{e}^{-p}$, and $\gamma_{e}$ is the 
Lorentz factor of the electrons), then the observed frequency should 
be above the synchrotron critical frequency for electrons with a minimum 
Lorentz factor ($\nu_{m}$) in the fast cooling phase or above the 
cooling frequency ($\nu_{c}$) in the slow cooling phase.  In this case, 
since there is no characteristic frequency above $\nu_{m}$ (in the fast 
cooling case) and $\nu_{c}$ (in the slow cooling case), it is reasonable 
to extrapolate upwards in energy.  The electron power-law index 
$2 \le p < 3$ is typical for both the prompt emission 
\citep[e.g.,][]{kaneko2006} and the afterglow of GRBs 
\citep[e.g.,][]{pk2002,yost2003}.  Thus, if we select the burst samples 
which have an observed photon index steeper than $2$ in the XRT data, 
the systematic error in the XRT-to-BAT extrapolation should be reduced.  

Following these arguments, we describe in this paper the XRT-to-BAT 
extrapolation to investigate the nature of the XRT steep decay
component.  As we mentioned, in principle the XRT-to-BAT extrapolation 
has similar issues to the BAT-to-XRT extrapolation.  
However, for certain bursts the XRT-to-BAT extrapolation could 
greatly reduce the systematic error.  More importantly, a different 
approach to the analysis may provide an alternate view of the problem.  

\section{Analysis}

X-ray light curves were produced from the XRT data for all 
{\it Swift} GRBs detected between June 2005 \footnote{Following updates 
to the on-board software which compensate for the uncontrolled 
temperature due to failure of the cooling control system 
\citep{kennea2005}, bright Earth effects and micrometeriod 
damage \citep{modes2}.} and September 2006.  54 GRBs with an early 
phase power-law decay index steeper than $-2$ 
($t_{0}$ taken as the BAT trigger time) were identified.  
From the spectral analysis of the X-ray data from these bursts in 
the 0.3 -- 10 keV band, we found that many bursts exhibited a significant 
difference in the best-fit spectral parameters between the early ($<$ 1000
seconds) data and the later ($>$ 1000 seconds) data.  This is 
likely due to the fact that $\ep$ is moving through XRT energy range early 
in the observation but one cannot measure this from the XRT data
alone because of the narrow energy band \citep[e.g.,][]{butler2007}.  To 
minimize the effects of the spectral evolution and also the absorption, 
only data above 2 keV was used in
these analysis.  We further refined the burst sample to those which satisfied the
following criteria: 1) Greater than four data bins containing at least 
20 counts in the 2--10 keV light curve during the early phase ($<$ 1000
seconds). This criteria reduced the number of bursts in the sample
significantly, due to much fewer counts being detected above 2 keV.  
2) No more than a single flare present in the early XRT light 
curve\footnote{Bursts exhibiting flares in the BAT were not excluded.}.  
3) The joint spectral analysis of the Photon Counting (PC) and 
Windowed Timing (WT) data must include a photon index of $2$ in the 90\% 
confidence interval.  13 GRBs satisfied the screening criteria.  Since the 2-10 keV 
joint spectral fit to the PC and WT data includes a best fit photon index steeper 
than $2$ for all the GRBs in the sample, $\ep$ should be below 2 keV.  
This will reduce the systematic error of the extrapolation because there should 
be no spectral evolution as a result of the shift of $\ep$ during the burst.  

As mentioned in \citet{zhang2006a}, the definition of the 
offset time ($t_{0}$) is critical when performing fits to 
the early phase light curve.  Traditionally, $t_{0}$ is defined as 
the trigger time of the GRB instrument; when the count rate exceeds 
some background level (rate trigger).  However, the definition of 
the BAT trigger time is different.  The BAT trigger time is 
the start time of the foreground time interval of the image from 
which the GRB is detected on-board.  Thus, to be comparable to a 
rate trigger time, we define $t_{0}$ as the start 
time of the prompt emission (start time of $t_{100}$ interval) for 
the whole sample.  

\subsection{BAT analysis}

The BAT analysis was performed using HEAsoft version 6.1.1 and CALDB 
version 2006-05-30.  The event-by-event data were used for these analysis.  
The non-linear energy correction for each event was applied by {\tt bateconvert}.  
The mask-weighting factors were calculated by {\tt batmaskwtevt} using 
the on-board position.  The detector enable and disable map was created by 
{\tt bathotpix} combining the enable and disable map generated by the flight 
software.  We created the BAT light curve by {\tt batbinevt} in the full energy range 
(15-350 keV\footnote{The coded mask is transparent to photons above 150 keV.  
Thus, photons above 150 keV are treated as background in the mask-weighted 
method.  The effective upper boundary is $\sim$ 150 keV.}) 
in 4 ms bin except for GRB 051109A (64 ms), GRB 060427 (1 s) and 
GRB 060923C (64 ms).  We used larger binning for these three GRBs 
because of the low signal-to-noise ratio of the emission.  The duration and 
the time intervals based on the Bayesian Block algorithm \citep{scargle1998} 
were calculated by {\tt battblocks}.  
The spectrum of each time interval was extracted by {\tt batbinevt}.  
The energy response file was created by {\tt batdrmgen}.  If the time 
interval was during the spacecraft slew, we updated the keywords in the 
spectral file related to the energy response process by {\tt batupdatephakw} 
and then created 
the energy response file for the time interval by {\tt batdrmgen}.  
We applied systematic error vectors to the spectrum using {\tt
batphasyserr}  prior to doing the spectral analysis.  The spectral
analysis was performed using Xspec 11.3.2.  

The energy flux in the 15--25 keV band was calculated for each time 
interval directly from the spectral fitting process.  The spectra 
from each time interval were fitted with a simple power-law model.  
According to the BAT GRB catalog (Angelini et al. in preparation), the 
detection threshold of the BAT in the 15--25 keV band is $\sim$ 10$^{-9}$ 
ergs cm$^{-2}$ s$^{-1}$.  Based on this result, the 15--25 keV flux 
was treated as an upper limit when the calculated 15--25 keV flux was 
less then 10$^{-9}$ ergs cm$^{-2}$ s$^{-1}$.  
The upper limit was estimated from using the event-by-event data from the Crab 
nebula on-axis observation collected on 2005 March 24 (observation ID: 
00050100016).  According to this observation, the BAT can detect the Crab 
nebula in the 15--25 keV band at 5 $\sigma$ in a one second exposure.  
Assuming that the BAT sensitivity scales as the square-root of the 
exposure time \citep{markwardt2005} and a canonical Crab flux of $5.3
\times 10^{-9}$ ergs cm$^{-2}$ s$^{-1}$ in the 15--25 keV band, 
we calculated the BAT 5 $\sigma$ 
upper limit in the 15--25 keV band from the following equation, 
\begin{equation}
F(15-25 {\rm keV})_{5\sigma} = \frac{5.3 \times 10^{-9}}{f_{pcode}}\, t_{exp}^{-0.5} \;({\rm ergs\,cm^{-2}\,s^{-1}}).
\end{equation}
Here $t_{exp}$ is the exposure time and $f_{pcode}$ is the partial 
coded fraction.  Since our estimation of the 5$\sigma$ upper limit is based on the on-axis Crab 
observation, $f_{pcode}$ will correct for a source observed in
the off-axis direction.  

For the time-averaged spectral analysis, we use the time interval from 
the emission start time to the emission end time ($t_{100}$ interval).  
When the spacecraft slew occurred during the time interval, we created the 
response matrices for each five second period, taking into account the position 
of the GRB in detector coordinates.  We then weighted these response 
matrices by the five second count rates and created the averaged 
response matrices.  Since the spacecraft slews about one degree per one 
second in response to a GRB trigger, we choose five second intervals to calculate 
the energy response for every five degrees.  

\subsection{XRT analysis}

The 13 bursts meeting the criteria outlined in section 2 were  
processed using the HEAsoft tools version 6.1.1 including version 2.5a 
of the {\it Swift} software.  The level 2 cleaned event files were  
produced from the {\tt xrtpipeline} task (version 10.4) using the  
standard screening criteria.  Version 8 of the response matrices in 
CALDB and the corresponding ancillary response produced from 
{\tt xrtmkarf} were used.  The standard grades \citep{burrows2005} 
were used in the analysis; grades 0-2 and 0-12 for WT and PC mode, respectively.

Both WT and PC mode data \citep{modes,modes2} from the first observation 
segment (000) were analyzed.  The source and background extraction
regions were nominally a 40-pixel square and an annulus with a 3 pixel 
inner radius and 25 pixel outer radius for WT and for PC mode,
respectively.  The extraction regions were modified for the piled-up
cases in WT mode in accordance with \citet{pileup2}, eliminating 
an inner square centered on the source.  It is well documented that 
pile-up in PC mode causes a redistribution of single pixel events to higher grades 
\citep{pileup}. In order to account for pile-up in PC mode, the grade 
distributions were analyzed for each burst during periods where the
count-rate exceeded 0.5 counts/sec.  For each burst, the
percentage of single pixel events were plotted versus the radius of 
the center of the annulus for increasing radii (0--8 pixels).  
The annulus with smallest radius at which the percentage of single 
pixel events became constant was used for the PC mode analysis.  Nominally, 
for a spectrum where the average energy is 1.5\ keV, 78\% of the events 
will be single pixel events \citep{moretti2004}.  

The hardness ratio ((2.0-10.0)keV/(0.5-2.0) keV) was examined for each
burst to ensure that there was no significant spectral evolution.  
An exposure map was created from {\tt xrtexpomap} to correct for the 
dead columns and hot pixels. {\tt xrtmkarf} was used to create the 
ancillary response files.  This includes corrections for losses in 
the wings of the point spread function and the center of the annulus, 
for the exposure and for vignetting. 

The spectrum file was binned with a minimum of 20 counts/bin 
in order for $\chi^2$ statistics to be valid for the spectral fitting.  
Xspec version 11.3.2 was used to perform a joint spectral fit of the 
WT and PC data from 2--10 keV using a simple power law model.  
The 15--25 keV normalization obtained from the $pegpwrlw$ model 
was used to extrapolate the XRT count rate into flux in the 
BAT 15--25 keV energy range.  Only 
light curve bins with greater than 90\% exposure were used in order to 
limit errors due to dead-time incurred by instrument mode switching. 
The light curve was binned to have greater than 20 counts/bin.

\subsection{Fitting a composite BAT and XRT light curve}

To investigate the connection between the prompt and the 
afterglow emission in the composite light curves, we first fit 
the XRT light curve, only using data where the hardness ratio 
showed no strong spectral evolution.   
We then fit both the BAT and XRT light curves jointly.  Both fits 
were performed using a power-law model with an offset time (PLO), 
\begin{equation}
F_{15-25 \rm keV} = K_{pow}\,(t-t_{0}^{pow})^{-\alpha}, 
\label{eq:curvature}
\end{equation}
where $t_{0}^{pow}$ is a offset time, $\alpha$ is a decay index, 
and $K_{pow}$ is the normalization, and with an exponential model (EXP), 
\begin{equation}
F_{15-25 \rm keV} = K_{exp}\,\exp(-\frac{t}{w}), 
\label{eq:exp}
\end{equation}
where $w$ is the decay constant and $K_{exp}$ is the normalization.  
For the XRT only fit, we fixed $t^{pow}_{0}$ to zero.
Finally, we fit the BAT and XRT light curves simultaneously using a 
combination of a power-law model with an exponential decay component (PLEXP), 
\begin{equation}
F_{15-25 \rm keV} =  K_{pow}\,(t-t_{0}^{pow})^{-\alpha} +
 K_{exp}\,\exp(-\frac{t}{w}).  
\label{eq:comb}
\end{equation}

For the fit to the XRT only light curve, the time interval 
was from the first XRT data point to the last data point before 
showing a deficit from the PLO model using an offset time of zero.  
For the joint BAT and XRT fit, 
the time interval was from the first BAT data point to the 
last XRT data point before showing a residual from the PLO model.  
Any other definitions used for the time interval are stated as a 
footnote in table \ref{tab:bat_xrt_lc_para}.  The values of the time 
intervals used in the light curve fitting are shown in the fourth 
column of table \ref{tab:bat_xrt_lc_para}.  
The best-fit model was selected based on the $\chi^{2}$ of the fit.  
However, because a PLO model will not fit the data before $t_{0}^{pow}$, 
the judgment between PLO and PLEXP is based on visual inspection 
as to whether the model fit both the BAT and XRT data simultaneously or not.  

\section{Results}

The left panels of figures \ref{fig:bat_xrt_lc1}--\ref{fig:bat_xrt_lc5} show the 
composite BAT (black open circles) and XRT (red open 
triangles) light curve in the 15-25 keV band overlaid with the best-fit 
light curve model.  The light curve models are PLO (eq. (\ref{eq:curvature})),  
EXP (eq. (\ref{eq:exp})), and PLEXP (eq. (\ref{eq:comb})) from top to bottom.  
In the bottom figure of PLEXP, both PLO and EXP  components are also shown as 
a dashed and a dash-dotted line, respectively.  
The best-fit parameters of the light curves are summarized in 
table \ref{tab:bat_xrt_lc_para}.  The best-fit of these models is 
labeled in blue.  The right panels of figures 
\ref{fig:bat_xrt_lc1}--\ref{fig:bat_xrt_lc5} from top to bottom 
show the BAT light curve in the 15-150 keV band, the BAT photon 
index based on the time-resolved spectral analysis, the XRT 2--10 keV 
count rate, and the XRT count rate ratio (2.0--10.0) keV/(0.5--2.0) keV.  
The best-fit spectral parameters based on the 2--10 keV joint fit to 
the XRT WT and PC mode data are summarized in table \ref{tab:xrt_spec_para}.

From the initial steep decay phase of the XRT light curve, it is difficult 
to distinguish between PLO and EXP from the XRT data alone.  Both models 
fit equally well for all of the bursts.  However, the difference and the 
importance of the individual components become clear when the BAT data
are included in the fit.  An EXP model fits well for GRB 050814, GRB 050915B,
GRB 060427 and GRB 060428B.  A PLO or a PLEXP model is not required for
these GRBs.  For GRB 060923C, a PLO is the model best represented by 
the composite light curve.  A PLEXP is the best model for the remaining 7 GRBs.  
The best-fit parameters which we used in the systematic study presented 
in this section, are shown in bold font in table \ref{tab:bat_xrt_lc_para}.  

\begin{figure}[pt]
\centerline{
\includegraphics[width=10cm,angle=-90]{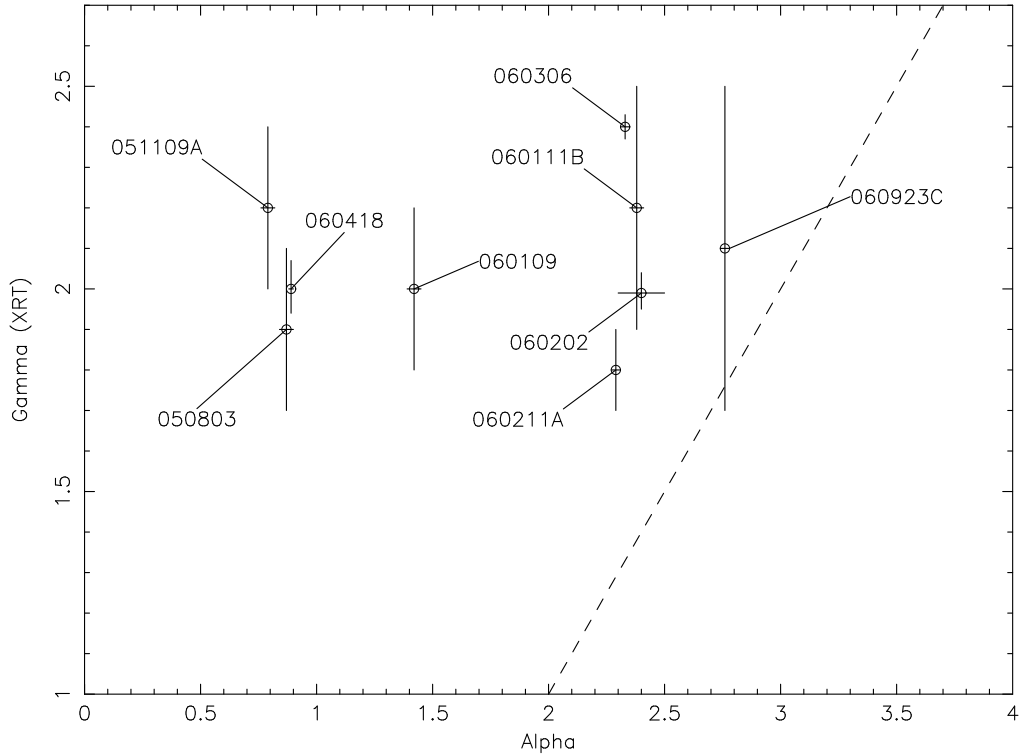}}
\caption{The relationship between the power-law decay index, $\alpha$, of the 
best-fit light curve with a PLO component and a photon index, $\Gamma_{XRT}$.  
The dashed line is the 
expected relationship from the curvature effect ($\alpha = 1+ \Gamma_{XRT}$).  
}
\label{fig:curvature}
\end{figure}

First, we investigated the possibility of the curvature effect for those 
GRBs which have a PLO component in the composite BAT and XRT light curve fit.  
According to the curvature effect \citep{fenimore1996,kumar2000,zhang2006a}, 
the relation between the decay index, $\alpha$, and the XRT photon
index, $\Gamma_{XRT}$, should be described by, $\alpha$ =
1+$\Gamma_{XRT}$, if the curvature effect is the cause of the XRT steep 
decay.  Figure \ref{fig:curvature} shows the correlation between 
$\alpha$ and $\Gamma_{XRT}$ for our sample.  The dashed line is the 
expected relationship from the curvature effect ($\alpha$ = 1+$\Gamma_{XRT}$).  
Although GRB 060923C may be consistent with the curvature effect, 
the majority of the bursts in the sample do not satisfy the expected relation.  
The inconsistency with the curvature effect could be due to neglecting 
the spectral evolution during the steep decay in our analysis of the XRT data.  
Looking at the time evolution of the count rate ratio of our sample, we 
find a hard to soft evolution from 0.6 to 0.5 and from 0.6 to 0.4 
for GRB 060202 and GRB 060211A.  These changes correspond to 
an evolution of the photon index from 1.5 to 1.7 and from 1.5 to 1.9, 
respectively, according to the calculation by the Xspec $fakeit$ command 
using the detector and the ancillary response files created for each source region.  
If this spectral evolution is taken into account, the steep decay could be 
consistent with the expectation of the curvature effect for GRB 060211A 
and GRB 060202.  However, we do not see a strong spectral 
evolution for the other GRBs with the exception of GRB 060418.  
Note that for GRB 060418, there is a strong evolution in the hardness 
ratio during the episode at $t_{0}$+150 s which may cause an error in 
the extrapolated flux.  

We find that most of the sample requires an EXP component to fit 
the BAT and XRT light curves simultaneously.  Therefore, we can conclude that 
some of the early steep decay observed by XRT is a continuation of the 
exponential decay tail of the prompt emission.  Interesting 
characteristics can be found for the bursts where a PLEXP 
model is the most representative model for the composite light curve.  
The dominant component of the fit to the XRT light curve 180 s 
after $t_{0}$ for GRB 060202 is an EXP.  For GRB 050803 
and GRB 060109 there is almost equal contribution from 
the EXP and PLO components in the initial XRT data .  Whereas, a PLO 
is the dominant component for GRB 051109A, GRB 060111B GRB 060211A, GRB 060306 
and GRB 060418B.  This result clearly demonstrates that the XRT steep decay 
could be composed of at least two different components.  
Without careful consideration of both the BAT and the XRT data simultaneously, 
it is not possible to distinguish between these two different
components.  It is important to note that \citet{obrien2006} also 
reached a similar conclusion; that the BAT and XRT composite light 
curve is composed of an exponential decay which relaxes to a power-law decay.  

\begin{figure}[pt]
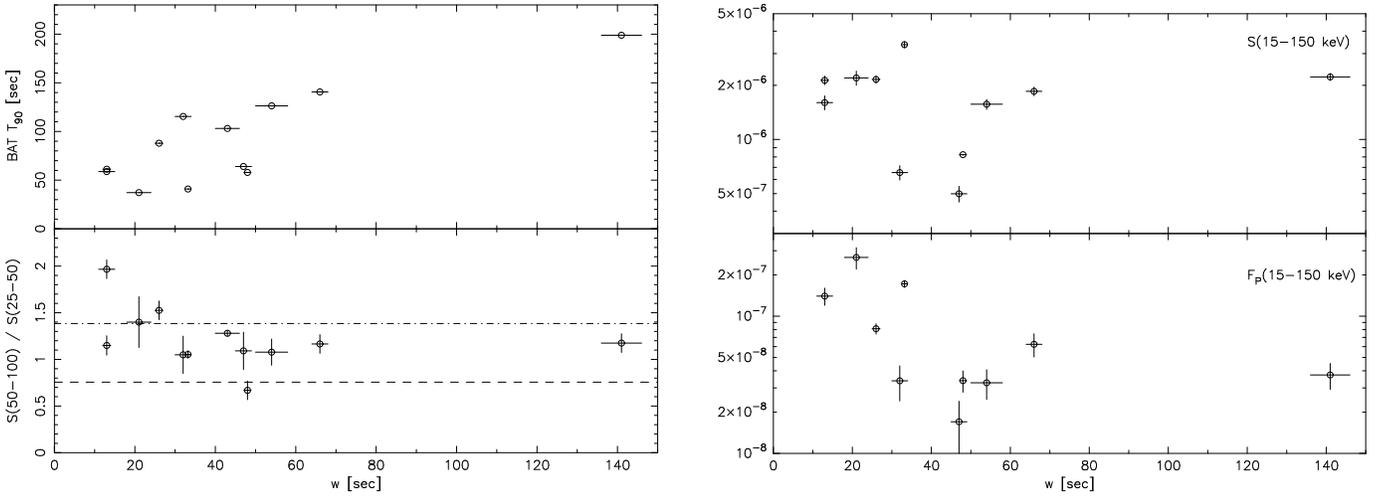

\centerline{
\includegraphics[width=6.5cm,angle=-90]{f8a.eps}
\hspace{0.5cm}
\includegraphics[width=6.5cm,angle=-90]{f8b.eps}}
\caption{The relationship between the decay constant, $w$, of the 
 exponential model and the BAT prompt emission properties.  Top left: 
 $w$ vs. BAT T$_{90}$, Bottom left: $w$ vs. the fluence ratio between the 
 50-100 keV and 25-50 keV band (the dotted and dash-dotted lines are 
the calculation assuming $\ep=30$ keV and $\ep=100$ keV, respectively,
 with a low and a high energy photon index of 1 and 2.5 in the Band function), 
Top right: $w$ vs. the fluence in the 15-150 keV band, 
Bottom right: $w$ vs. the 1-s peak flux in the 15-150 keV band.}
\label{fig:cor_w}
\end{figure}

For the bursts that exhibit an EXP component, we investigated the 
correlation between the exponential decay index, $w$, and the prompt emission 
properties derived from the BAT data (table \ref{tab:bat_grb_para}).  
The results are summarized in figure \ref{fig:cor_w}.  
No correlation is found for the properties of the prompt emission 
except between $w$ and the BAT T$_{90}$ which is expected because both 
parameters are related to the duration of the bursts.

\begin{figure}
\centerline{
\includegraphics[width=12cm,angle=-90]{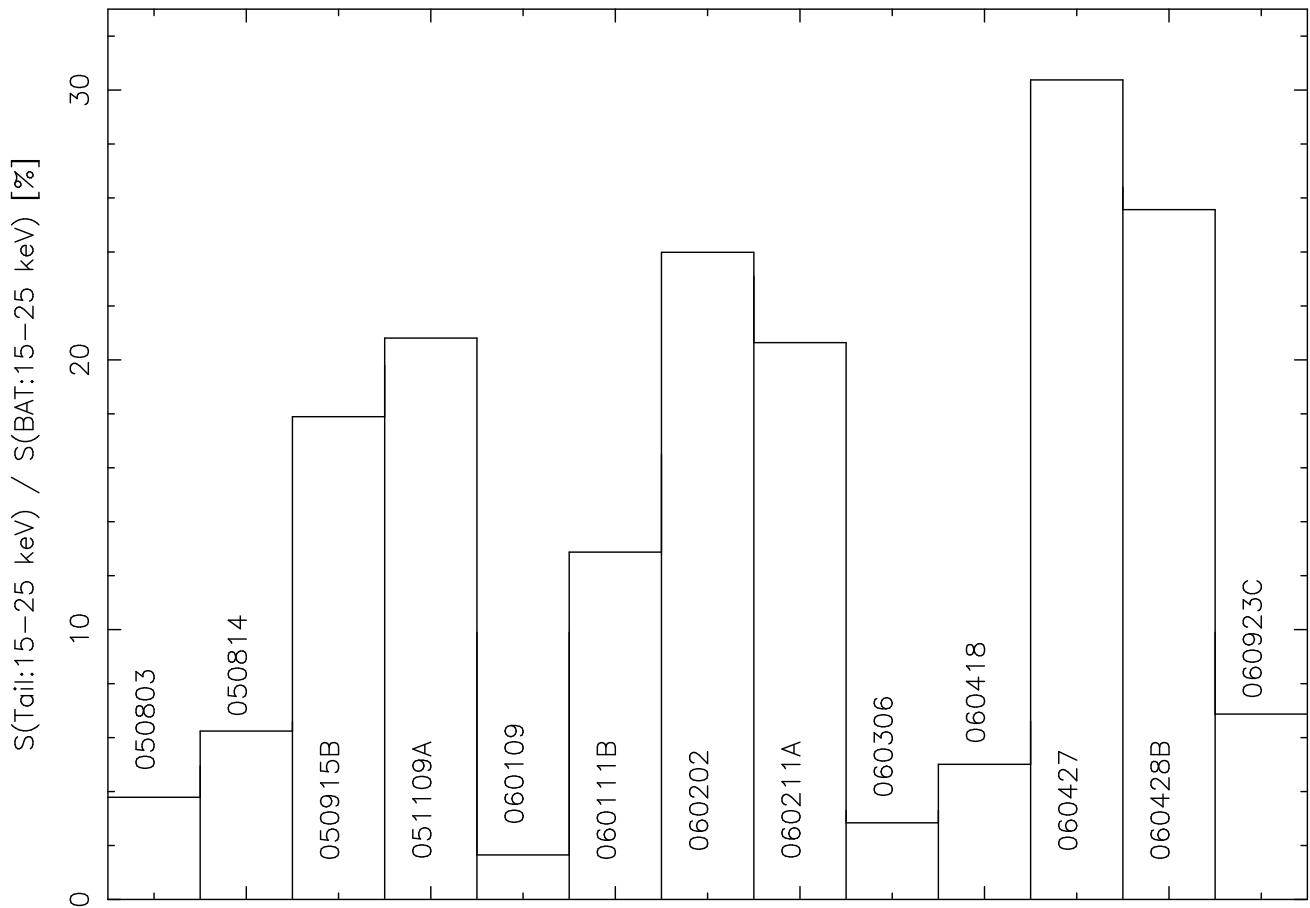}}
\caption{A histogram of the ratio, in percent, between the fluence 
accumulated from the end of the emission as detected by the BAT to 1000 s  
after $t_{0}$ (S(Tail:15-25 keV)) and the fluence recorded by the BAT 
(S(BAT:15-25 keV)).}
\label{fig:BAT_XRT_fluence_hist}
\end{figure}

Based on our study, there is a strong indication that the steep decay 
component observed by the XRT is part of the prompt emission \citep[e.g., also][]
{nousek2006,obrien2006}.  Thus, we calculate the fluence which may be 
below the sensitivity limit of the BAT.  This fluence was calculated by 
accumulating the flux from the best-fit composite BAT and XRT light 
curve model from the end of the emission as detected by the BAT to 1000 
seconds after $t_{0}$.  Figure \ref{fig:BAT_XRT_fluence_hist} shows 
the ratio of the percentage of the fluence in the tail emission which 
is below the BAT sensitivity limit and the fluence recorded by the BAT.  
For 7 out of the 13 GRBs in the sample, the fluence of the tail component 
is less than 15\% of the fluence recorded by BAT.  However, more than 15\%
of the fluence may be radiated below the BAT sensitivity limit for 
GRB 050915B, GRB 051109A, GRB 060202, GRB 060211A, GRB 060427, and 
GRB 060428B.  This result gives rise to the question as to
whether the fluence measured by the $\gamma$-ray instrument reflects 
the true fluence of the prompt emission.  

\section{Discussion}

We have presented the BAT and XRT composite light curves, derived 
extrapolating the XRT 2-10 keV flux up to the BAT 15--25 keV energy 
range for GRBs which have a steep decay component in the initial XRT 
light curve.  Based on the simultaneous fit of 
both the BAT and XRT light curves, we have confirmed the existence of an EXP 
component which smoothly connects the BAT prompt emission to the 
XRT steep decay for several GRBs.  We have also found that the XRT  
steep decay for some of the bursts can be fit well by a PLEXP 
model.  In the following sections, we discuss the possible origins of the 
PLO and EXP components.

\subsection{Origin of the PLO component}

\begin{figure}[pt]
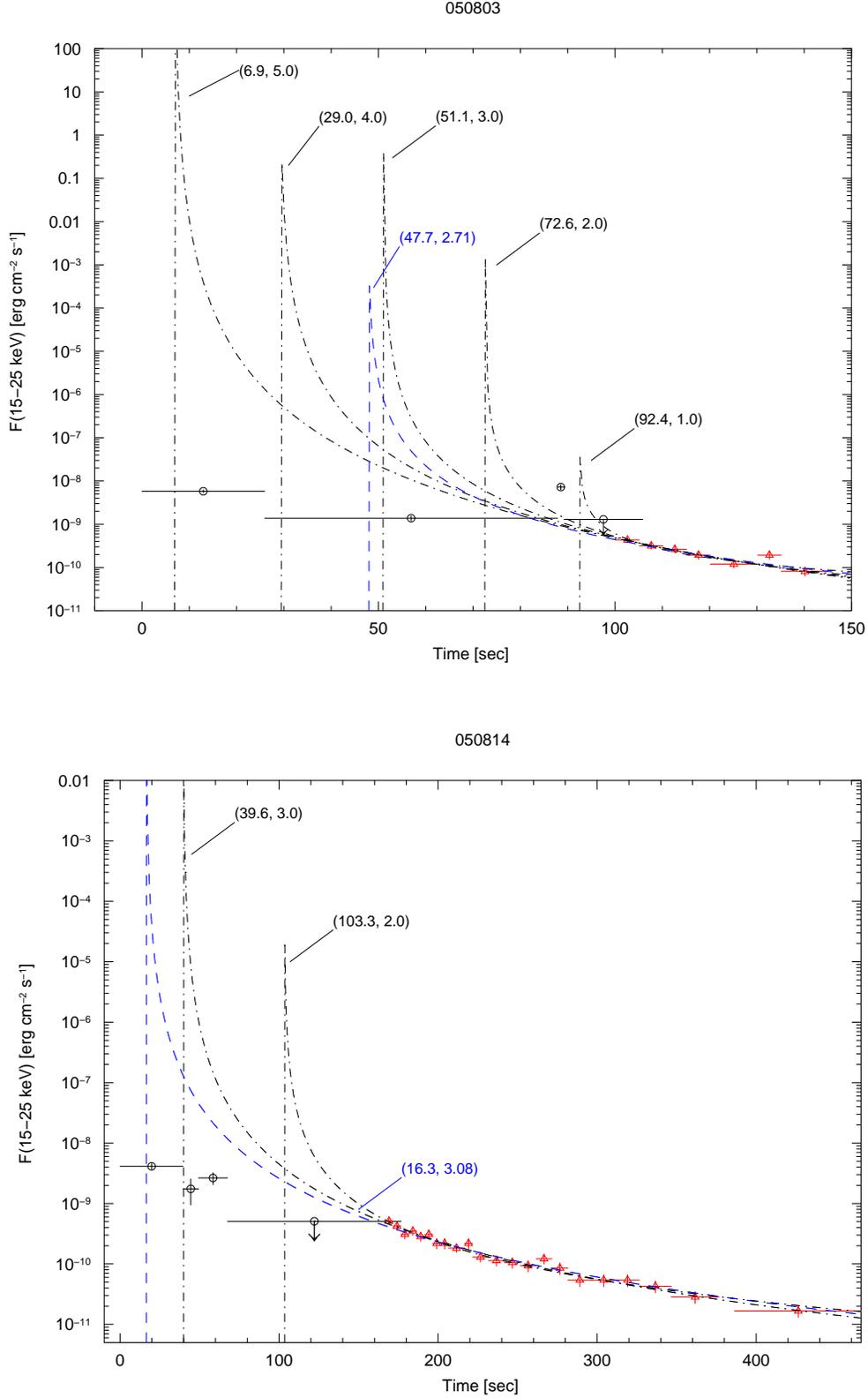

\centerline{
\includegraphics[width=10cm,angle=-90]{f10a.eps}}
\vspace{1cm}
\centerline{
\includegraphics[width=10cm,angle=-90]{f10b.eps}}
\caption{The BAT and XRT composite light curves of GRB 050803 (top) and 
GRB 050814 (bottom) overlaid with the best-fit PLO model for different 
choices of $\alpha$ and $t_{0}^{pow}$.  The labels in the parentheses are 
($t_{0}^{pow}$, $\alpha$).  The model which represents \citet{liang2006} 
is shown with a blue dashed line.}
\label{fig:linag_prob}
\end{figure}

A PLO component most likely originates from an internal shock 
\citep[so called, curvature radiation or high-latitude emission
associated with the last bright spike.][] {kumar2000,nousek2006,zhang2006a,yamazaki2006}.  
Our results support this idea, because for most bursts the 
XRT steep decay component smoothly connects with the last 
bright episode detected by the BAT (e.g. GRB 050803).  The instantaneous 
emission from a uniform jet produces a decay index of 
$\alpha = 1 + \Gamma$ \citep{kumar2000}.  This formula was examined 
using the power-law decay index derived from the PLO model and the 
photon index based on the joint WT and PC spectral analysis of our 
sample.  For the majority of the sample, we find that the decay index 
is not consistent with the formula.  One of the possible reasons for 
this inconsistency could be the spectral evolution during the 
early XRT observation in some bursts \citep{zhangbb2006}.  
However, as  discussed in section 3,  the inconsistency cannot always 
be associated with spectral evolution.  Another possible reason is 
the choice of the time-zero \citep{zhang2006a,yamazaki2006}. 
\citet{liang2006} investigated the curvature effect as an origin 
of the XRT steep decay using the data set of \citet{obrien2006}.  
They made the assumption that the XRT steep decay component is due 
to the curvature effect and investigated whether the time-zero is 
consistent with the beginning of the bright episode.  They concluded 
that for most of the sample, the time-zero was consistent with this picture.   
The main issue with their approach is that for the fixed power-law decay 
index, as expected from the formula $\alpha = 1+\Gamma$,  
by shifting the offset time it is possible to fit the decay index 
to most early XRT data.  This is because in their approach, the fitting 
parameters are not only the offset time but also the normalization, 
allowing extra freedom in the fit.  Here, we demonstrate this problem using 
GRB 050803 and GRB 050814.   The dash-dotted lines in figure 
\ref{fig:linag_prob} show the best fit PLO model by changing 
$\alpha$ from 1 to 5 for GRB 050803 and from 2 to 3 for GRB 050814, 
and varying $t_{0}^{pow}$.  Only the XRT data, shown in red triangles, 
are used in the fitting process as in \citet{liang2006}.  As seen in the figures, 
the choice of $\alpha$ and $t_{0}^{pow}$ is not unique if one only try to fit 
the XRT data.  Moreover, as it is clearly demonstrated in the case of GRB 050814, 
even if $t_{0}^{pow}$ is chosen as the start time of the GRB pulse, the intensity 
of the pulse expected from the model is an order of magnitude brighter than 
the data.  Therefore, if a bright episode in 
the BAT data, which could contribute to the steep decay component, is 
simultaneously fit with the XRT data, as in our approach, both the 
offset time and the decay index will be uniquely constrained by the data.  
It may be difficult to test the curvature effect or the relation 
$\alpha = 1+\Gamma$ definitely without fitting the XRT and BAT 
data simultaneously.  The third possibility is to abandon the assumption
of the uniform jet emission.  Our results may suggest 
that the structure of the jet is much more complex than a uniform jet
\citep[e.g.][]{yamazaki2006}.  

\subsection{Origin of EXP component}

Since it is difficult to explicitly state the origin of an EXP component, 
we will discuss the possibilities of both an internal shock 
and an external shock as the origin of an EXP component.  

\subsubsection{External shock scenario for EXP component}

One interpretation of an EXP component is the presence of the 
external shock emission during the prompt phase.  \citet{fenimore1999} 
studied the case of the co-existence of the emission from the external shock 
(deceleration of the initial shell) during the emission from the internal 
shocks.  They showed that the smooth long lasting soft emission which arose 
from an external shock could overlay the light curve of the 
prompt emission.  Furthermore, \citet{fenimore1999} showed that the efficiency 
of converting the bulk energy to radiation is 85\% in this case,   
whereas, internal shocks without deceleration only convert 
about 1\% \citep{kumar1999,panaitescu1999}.  \citet{zhang2006b} 
calculated the radiation efficiency using the {\it Swift} X-ray afterglow 
data and show that about half of the GRBs have an 
efficiency $\gtrsim$ 1\% which provides a challenge for producing the prompt 
emission from internal shocks alone.  Furthermore, according to the 
optical observation by Rapid Telescopes for Optical Response (RAPTOR) 
during the prompt emission of GRB 050820A, 
smoothly decaying emission which does not correlate with the prompt 
spikes was found \citep{raptor_050820a}.  If this emission is from 
the external shock, we might be observing the deceleration of the 
outflow during the prompt phase.  

Following the argument of \citet{fenimore1996} and \citet{fenimore1999}, 
we estimate the bulk Lorentz factor of our sample, assuming 
an EXP component is purely due to external shock emission.  
Let us assume that an external shock starts its emission at the radius, $R_{0}$.  
The external shock will be decelerated because of sweeping up the 
inter-stellar medium (ISM).  The total energy of the central engine, 
$E_{0}$, can be expressed as, 
\begin{equation}
E_{0} = (4 \pi / 3) R_{0}^{3} n_{\rm ISM} m_{p}c^{2} \gamma_{0}^{2}
\label{eq:e0}
\end{equation}
where $n_{\rm ISM}$ is a density of the ISM, $m_{p}$ is the proton mass, c is 
the speed of light, and $\gamma_{0}$ is the bulk Lorentz factor at $R_{0}$.  
The duration of the emission ($\Delta T$) is determined by the radial time 
scale \citep{piran1999} which is the difference in the arrival time of the 
photons emitted between $R_{0}$ and $aR_{0}$ ($a>1$)  as measured by the observer, 
\begin{displaymath}
\Delta T = [(a^{4}-1)/4] (R_{0}/2 \gamma_{0}^{2}c).  
\end{displaymath}
Therefore, the bulk Lorentz factor can be expressed as, 
\begin{equation}
\gamma_{0} = (32 \pi m_{p} c^{2} / 3)^{-1/8} [(a^{4}-1)/4]^{3/8}
 (E_{0}/n_{ISM})^{1/8} (c \Delta T)^{-3/8}.  
\label{eq:gamma0}
\end{equation}
The relationship between $\Delta T$ and the full width at half maximum 
(FWHM) of the pulse ($T_{1/2}$) is described by, 0.22 $\Delta T$ = $T_{1/2}$, which is 
valid for a pulse shape of a fast rise, exponential decay (FRED) 
\citep{fenimore1996}.   Thus, once we know the redshift, since $T_{1/2}$ can be 
estimated from the best-fit parameters of an EXP component, we can calculate 
the bulk Lorentz factor as a function of $E_{0}$/$n_{\rm ISM}$.  In the following 
arguments, we use a typical value of 2 for the parameter ``$a$'' 
(in the $a=2$ case, $\Delta T$ is the radial time scale from $R_{0}$ to 2$R_{0}$).  

For bursts with unknown redshifts (all bursts in the sample except 
GRB 050814, GRB 051109A and GRB 060418),  we used the mean redshift of
2.4 obtained from the {\it Swift} long GRBs\footnote{http://heasarc.gsfc.nasa.gov/docs/swift/archive/grb\_table/}.  
We can derive a reasonable range of
$\gamma_{0}$ from 143 to 350 and $R_{0}$ of $\sim$ 10$^{16}$ cm 
assuming $E_{0}/n_{\rm ISM} = 1 \times 10^{52}$ erg cm$^{3}$ 
for the 12 GRBs in our sample which have an EXP component in the best
fit model (table \ref{tab:bat_grb_para}).  According 
to the calculation of $\gamma_{0}$ by the Rapid Eye Mount (REM) telescope 
using the peak time of the early afterglow data, $\gamma_{0}$ is about 
400 for GRB 060418 and GRB 060607A \citep{molinari2006}.  
Their $\gamma_{0}$ value agrees within a factor of two of our estimates.  
For GRB 060418, $E_{0}$ will be $\sim$ $1 \times 10^{54}$ 
ergs if we assume $\gamma_{0}$ of 400 as derived from the REM observation and 
$n_{\rm ISM}$ of 1 cm$^{-3}$.  $E_{0}$ of $\sim$ $10^{54}$ ergs is also 
a typical value according to the calculation of \citet{zhang2006b} 
using X-ray afterglow data observed by XRT.  

It would be interesting to look for a correlation between 
$\gamma_{0}$ and $\ep$ in the GRB rest frame 
($\eps$) although it is difficult to calculate for our sample because 
we do not have measurements of both $\ep$ and redshift.  
One of the advantages of using the {\it Swift} sample for this 
study is that soft GRBs, so called X-ray Flashes (XRFs), are included 
in the sample because of the relatively softer energy response of 
BAT compared to that of BATSE.  In the dirty fireball model 
\citep{dermer1999}, $\eps$ has a strong dependency on the bulk Lorentz 
factor ($\eps \propto \gamma_{0}^{4}$).  The unified jet models for 
XRFs and GRBs, such as the structured-jet model \citep{rossi2002}, 
and the variable jet opening angle model \citep{lamb2006,tim2006}, 
expect a positive correlation between $\eps$ and the bulk Lorentz 
factor.  In the off-axis jet model \citep{yamazaki2004}, 
no correlation is expected between $\gamma_{0}$ and $\eps$ because the Doppler factor 
will change as a function of a viewing angle but not $\gamma_{0}$.  
Another interesting theoretical model to discuss is 
the case of a very high Lorentz factor \citep{mochkovitch2003,barraud2005}.  
According to this model, XRFs can be produced in a condition with 
a very high $\gamma_{0}$ (so called ``clean fireball''), while 
classical GRBs have a moderate $\gamma_{0}$.  In this case, we would expect 
a negative correlation between $\eps$ and the bulk Lorentz factor.  
Some additional effort, for example, to estimate $\ep$ from the {\it Swift} BAT data 
(Sakamoto et al. in preparation) and to estimate a redshift from the {\it Swift} data 
\citep[e.g.,][]{grupe2006} is encouraged in order to discuss 
the correlation between $\gamma_{0}$ and $\eps$ and the origin of XRFs.  

We can derive another constraint on $\gamma_0$ which is independent of 
the previous discussion.  Our observational results suggest that the photons
originating from an internal shock via the curvature effect (PLO component) 
and photons from an external shock (EXP component) 
arrive at the observer almost simultaneously.  The observed time of the 
photon from an internal shock, $T_{int}$, can be expressed as, 
\begin{displaymath}
T_{int} \sim (R_{int}/2c\gamma_{int}) [1+(\gamma_{int}\theta)^{2}], 
\end{displaymath}
where $R_{int}$ is the radius where an internal shock emits,
$\gamma_{int}$ is the bulk Lorentz factor at $R_{int}$, and $\theta$ 
is the jet opening half-angle.  On the other hand, the observed time of 
an external shock emission, $T_{ext}$, is expressed as, 
\begin{displaymath}
T_{ext} \sim R_{ext}/(4c\gamma_{ext}^2), 
\end{displaymath}
where $R_{ext}$ is the radius where an external shock emits and $\gamma_{ext}$ is a 
bulk Lorentz factor at $R_{ext}$ \citep{sari1998a}.  If we assume 
$T_{int} \sim T_{ext}$ and also $\gamma_{int} \sim \gamma_{ext} \sim \gamma_{0}$, 
we have, 
\begin{displaymath}
2R_{int}[1+(\gamma_{0}\theta)^{2}] \sim R_{ext}.  
\end{displaymath}
$\eps$ can be 
written as a function of $\theta$ and $\gamma_{0}$, 
\begin{displaymath}
\eps(\theta) \sim (2\gamma_{0} h \nu_{0}^{\prime})[1+(\gamma_{0} \theta)^{2}] \sim 
[\eps (\theta = 0)] / (1 + (\gamma_{0} \theta)^{2}), 
\end{displaymath}
where $\eps (\theta = 0)$ 
is $\ep$ observed by the on-axis observer.  In this case, $\eps (\theta = 0)$ 
corresponds to the observed $\ep$ multiplied by (1+z).  
Therefore, the relationship 
between $R_{int}$, $R_{ext}$ and $\ep$ is given by, 
\begin{displaymath}
R_{ext}/2 R_{int} \sim [\ep (\theta = 0)]/\ep(\theta).  
\end{displaymath}
Since the observed photon index of the XRT steep decay emission of 
our sample is $\sim 2$, which suggests that the observation of the 
spectrum is above $\ep$, it is reasonable to assume that the upper 
limit of $\ep(\theta)$ is in the few keV range.  Hence, the condition will be, 
\begin{displaymath}
R_{ext}/2 R_{int} > [\ep (\theta = 0)]/\ep(\theta^{\prime}) 
\end{displaymath}
where $\ep(\theta^{\prime}$) is the upper limit of a few keV.  
If we use the angular spreading time ($\Delta T_{ang}$) \citep{piran1999} 
as the time scale of an internal shock, then, 
\begin{displaymath}
R_{int} \sim 2 c \gamma_{0} \Delta T_{ang}.  
\end{displaymath}
From Eq.~(\ref{eq:e0}), the radius of an external shock can be expressed as 
\begin{displaymath}
R_{ext} = (4 \pi m_{p}c^{2}/3)^{-1/3} (E_{0}/n_{ISM})^{1/3} \gamma_{0}^{-2/3}.  
\end{displaymath}
Thus, $\gamma_{0}$ can be written as, 
\begin{equation}
\gamma_{0} > (1/4)^{3/8} (4 \pi m_{p}c^{2}/3)^{-1/8} (E_{0}/n_{ISM})^{1/8} (c \Delta T_{ang})^{-3/8} 
[\ep (\theta = 0)/\ep(\theta^{\prime})]^{-3/8}.  
\end{equation}

In the case of GRB 050803, $\Delta T_{ang}$ can be derived as the
duration of the last spike (the pulse at $t_{0}$+90 s), $\Delta T_{ang}$ 
$\sim$ 3 s, if one takes into account the time dilation effect.  
Although it is not possible to extract the information about the $\ep$ 
of this pulse from the BAT data, it is reasonable to assume $\ep$ $>$ 
100 keV since the photon index from a simple power-law fit to the BAT 
data is $1.2 \pm 0.2$ which is close to the low energy photon index 
of the typical GRB spectrum ($\ep$ should be around or 
above the BAT upper energy range of 150 keV).  Therefore, we use 
$\ep (\theta = 0)/\ep(\theta^{\prime}) > 100$.  The 
lower limit of $\gamma_{0}$ of GRB 050803 is estimated to be $<$ 50 
assuming a redshift of 2.4 (the mean redshift of the {\it Swift} 
long GRBs) and $E_{0}/n_{ISM}$ of $1 \times 10^{52}$ ergs cm$^{3}$.  
Applying the same assumptions for the GRB spectral parameters 
($\ep (\theta = 0)/\ep(\theta^{\prime}) > 100$), we estimate 
a lower limit of $<$ 60 for GRB 060418 
using the measured redshift 
of 1.489, $\Delta T_{ang}$ of 2 s (the pulse at $t_{0}$+50 s), and 
$E_{0}/n_{ISM}$ of $1 \times 10^{52}$ ergs cm$^{3}$.  The estimated 
lower limit of $\gamma_{0}$ for GRB 060418 is not contradicted by the value based 
on the REM observation.  
In summary, for the EXP component, one can provide a reasonable bulk Lorentz 
factor, $\gamma_{0}$, within the external shock scenario, which is consistent 
with other measurements.  

\subsubsection{Internal shock scenario for EXP component}

As we discussed in \S 4.1, a simple interpretation based on a uniform
jet model could contradict the prediction of the curvature effect.  
Here we discuss an internal shock scenario for both EXP and PLO
components based on an inhomogeneous jet model.  
\citet{yamazaki2006} investigated the GRB prompt emission 100-1000 
seconds after the GRB trigger within the frame work of a multiple sub-shell model.  
According to their study, despite an angular inhomogeneity of the jet, 
the tail emission has a monotonic decay which resembles the XRT steep decay.  
In this context, if the jet has a core in which the emission energy is 
densely confined compared with the outer region, the PLO decay 
component arising from an on-beam sub-shell may be overlaid by 
the off-beam core emission which causes the EXP decaying component.   
\citet{takami2006} further extended their study, and in order to 
investigate the unknown jet structure, they proposed unique 
definitions of the decay index derived by unique definitions of 
the time-zero and of the fitting interval of the observed light curve.
They found that the decay index in their definitions should have 
a wide scatter in the case of a power-law like structured jet. 
Here, we calculated the decay index using our BAT and XRT composite 
light curve based on the definitions of \citet{takami2006}.  Because of the 
difficulty of using exactly the same definition of time-zero 
as proposed in \citet{takami2006} ($T_{*}$ in their paper), we define 
the end of the BAT emission 
as $T_{*}$.  We can fit our light curves using the proposed fitting interval 
for five GRBs in our sample ($\alpha_{tail}$ as a decay index, $\chi^{2}$ and 
d.o.f. of the fits are shown in the last two columns of table \ref{tab:bat_grb_para}).  
Our results, based on a very small sample, show that the decay index 
ranges from 0.6 to 3.0.  However, both the size of the sample and 
the number of data points included in the light 
curve fit are very small because of using XRT data above 2 keV.  
There is an additional issue further reducing the sample; the 
appearance of a shallow decay in the XRT data in the fitting interval.  
Unfortunately, it is hard to conclude the structure of a jet with our 
limited sample.  Once the GRB sample suitable for fitting a light curve 
with the unique definitions of \citet{takami2006} can be increased, 
we may be able to draw a conclusion about the jet structure of 
GRBs using the XRT steep decay component.  

\section{Summary}

In this paper, we presented a systematic study of the steep decay 
emission observed by the XRT.  We constructed composite light curves 
in the 15--25 keV band extrapolating the XRT data (2--10 keV) up to 
the BAT energy range (15--25 keV).  Based on the simultaneous fitting 
of the BAT and XRT data, we confirmed the existence of an EXP component 
for the majority of the bursts in the sample.  We found that for the 
PLO component, the majority of the GRBs in our sample are inconsistent 
with the relationship of the curvature effect, $\alpha = \Gamma + 1$, 
which is only valid in the case of the uniform jet.  We also found that more 
than 15\% of the prompt fluence may be radiated below the BAT 
sensitivity limit for half of our sample.  We argue that 
the EXP component could be the emission from the external shock 
which may indicate the deceleration of the initial shell by ISM during 
the prompt phase.  We discuss the case of the prompt tail emission 
from the structured jet as an origin of the XRT steep decay but the
sample is too small for a solid conclusion.  

\acknowledgements
We would like to thank A. P. Beardmore, G. Chincarini, C. Guidorzi, 
P. T. O'Brien and K. L. Page for valuable comments.  We also would like 
to thank the anonymous referee for comments and suggestions that materially 
improved the paper.  
This research was performed while T.~S. held a NASA Postdoctoral Program 
administered by Oak Ridge Associated Universities at NASA Goddard Space 
Flight Center.  R.~Y. was supported in part by Grants-in-Aid for Scientific Research
of the Japanese Ministry of Education, Culture, Sports, Science,
and Technology 18740153.  The material of the paper has been improved by
the discussions during the workshop ``Implications of {\it Swift's}
Discoveries about Gamma-Ray Bursts'' at the Aspen Center for Physics.

\clearpage

\newpage
\begin{deluxetable}{ccccccccccc}
\tabletypesize{\scriptsize}
\rotate
\tablecaption{Parameters of the light curve fits. Errors quoted at 68\%
 confidence level.  See text for details (section 2.3 and 3).\label{tab:bat_xrt_lc_para}}
\tablewidth{0pt}
\tablehead{
\colhead{GRB} & 
\colhead{$t_{0}$} &
\colhead{Data$^{\ddagger}$} & 
\colhead{Fitting} & 
\multicolumn{4}{c}{Power-law} &
\multicolumn{3}{c}{Exponential}\\\cline{5-11}
\colhead{} &
\colhead{UT} &
\colhead{} &
\colhead{[s]} &
\colhead{t$_{0}^{pow}$} & 
\colhead{$\alpha$} &
\colhead{$K_{\rm pow}$} & 
\colhead{$\chi^{2}$/dof} & 
\colhead{$w$} & 
\colhead{$K_{\rm exp}$} & 
\colhead{$\chi^{2}$/dof}
}
\startdata
050803 & 2005-08-03 & XRT & 100--147 & 0$^{\dagger}$ & $5.3 \pm 0.7$ &
  17.4 & 6.6 / 5 & $22 \pm 3$ & $3.8_{-1.9}^{+4.0} \times 10^{-8}$ & 7.3 / 5 \\
       &    19:14:59.3                  & PL/EX & 0--406 & $87.2_{-0.4}^{+0.3}$ &
 $1.18 \pm 0.07$ & $(1.1 \pm 0.3) \times 10^{-8}$ & 14.4 / 10 & $32.3 \pm 0.9$ &
 $(8.4 \pm 0.7) \times 10^{-9}$ & 122.2 / 13\\
       &                         & PLE & 0--406 & {\boldmath $88.3_{-0.05}^{+0.03}$} 
& {\boldmath $0.87 \pm 0.03$} & {\boldmath $2.0 \times
 10^{-9}$$^{\dagger}$} & {\bf --} & {\boldmath $26 \pm 1$} & {\boldmath
 $1.0 \times 10^{-8}$$^{\dagger}$} & {\bf 22.0 / 12} \\\hline
050814 & 2005-08-14 & XRT & 167--466 & 0$^{\dagger}$ & $3.6 \pm 0.3$ & 
 $4.1_{-2.4}^{+6.2} \times 10^{-2}$ & 11.7 / 20 & $71 \pm 4$ & $4.0 \times
 10^{-9}$ & 23.7 / 20 \\
       &   11:38:55.4  & PL/EX & 0--466 & 70.3 & 2.5 & $5.3 \times
 10^{-5}$ & 10.3 / 21 & {\boldmath $66 \pm 2$} & {\boldmath $(5.0 \pm
 0.5) \times 10^{-9}$} & {\bf 27.7 / 23}\\
       &               & PLE & 0--466 & 44.5 & $1.6$ & $5.1 \times
 10^{-8}$$^{\dagger}$ & -- & $62 \pm 1$ & $5.8 \times 10^{-9}$$^{\dagger}$ & 24.2 / 22 \\\hline
050915B & 2005-09-15 & XRT & 158--228 & 0$^{\dagger}$ & $5.3
 \pm 0.8$ & $1.3 \times 10^{2}$ & 10.7 / 5 & $36_{-5}^{+6}$ & 
 $2.1_{-1.2}^{+2.7} \times 10^{-8}$ & 11.7 / 5 \\
        & 21:22:56.6 & PL/EX & 7--228 & 21.3 & $2.1$ & $4.0 \times
 10^{-6}$ & 27.0 / 7 & {\boldmath $33.2 \pm 0.6$} & {\boldmath $(3.0 \pm
 0.1) \times 10^{-8}$} & {\bf 27.6 / 8}\\
        &                         & PLE & 7--228 & $-29_{-10}^{+11}$ &
 $3.8^{+0.3}_{-0.2}$ & $2.0 \times 10^{-2}$$^{\dagger}$ & -- & 
 $35_{-4}^{+2}$ & $1.7 \times 10^{-8}$$^{\dagger}$ & 16.4 / 7\\\hline
051109A & 2005-11-09 & XRT & 131--196 & 0$^{\dagger}$ & $2.2 \pm 0.7$ 
 & $3.2 \times 10^{-6}$ & 0.4 / 4 & $74_{-19}^{+36}$ &
 $5.0_{-2.5}^{+5.2} \times 10^{-10}$ & 0.3 / 10 \\
        & 01:12:17.6 & PL/EX & 0--196 & $1.9$ & -1.6 & $2.2 \times 10^{-7}$ & 7.8 / 6
 & $28 \pm 1$ & $(1.5 \pm 0.2) \times 10^{-8}$ & 50.8 / 8\\
        &  & PLE & 0--196 & {\boldmath $4.78 \pm 0.02$} & {\boldmath $0.79 \pm 0.03$} 
& {\boldmath $2.5 \times 10^{-9}$$^{\dagger}$} & {\bf --} & {\boldmath $21 \pm 3$} 
& {\boldmath $1.5 \times 10^{-8}$$^{\dagger}$} & {\bf 2.8 / 7}\\\hline
060109 & 2006-01-09 & XRT & 110--200 & 0$^{\dagger}$ & $4.3 \pm 0.4$ & 
$2.6 \times 10^{-1}$ & 7.3 / 8 & $33 \pm 3$ &
 $8.6_{-2.7}^{+4.0} \times 10^{-9}$ & 8.5 / 8\\
       & 16:54:41.2 & PL/EX & 0--862 & $88 \pm 1$ & $1.6 \pm 0.1$ &
 $7.1_{-2.2}^{+3.7} \times 10^{-8}$ & 11.1 / 11 & $42 \pm 2$ & $(3.6 \pm
 0.4) \times 10^{-9}$ & 109.1 / 12\\
       &                         & PLE & 0--862 & {\bf 90 $\pm$ 0.3} &
 {\boldmath $1.42 \pm 0.03$} & {\boldmath $1.8 \times
 10^{-8}$$^{\dagger}$} & {\bf --} & {\boldmath $32 \pm 2$} & {\boldmath $3.5 \times
 10^{-9}$$^{\dagger}$} & {\bf 8.4 / 11} \\\hline
060111B & 2006-01-11 & XRT & 89--149 & 0$^{\dagger}$  &$2.8
 \pm 0.7$ & $2.1 \times 10^{-5}$ & 2.6 / 3 & $43_{-9}^{+14}$ &
 $6.5_{-3.3}^{+6.2} \times 10^{-10}$ & 3.2 / 3\\
        & 20:15:41.2 & PL/EX & 0--149 & 50 & $2.4$ & $9.3 \times
 10^{-7}$ & 13.0 / 6 & $23 \pm 1$ & $(5.7 \pm 0.7) \times 10^{-9}$ &
 133.3 / 7\\
        &                         & PLE & 0--149 & {\boldmath $50.1 \pm 0.3$} &
{\boldmath $2.38 \pm 0.03$} & {\boldmath $7.5 \times 10^{-7}$$^{\dagger}$} & {\bf
 --} & {\boldmath $13 \pm 2$} & {\boldmath $7.2 \times 10^{-9}$$^{\dagger}$} & {\bf 14.5 / 6}\\\hline
060202 & 2006-02-02 & XRT & 250--350 & 0$^{\dagger}$ & $2.2 \pm 0.3$ & 
 $8.7 \times 10^{-5}$ & 14.8 / 18 & $138_{-16}^{+20}$ &
 $3.5_{-0.8}^{+1.1} \times 10^{-9}$ & 15.2 / 18\\
       & 08:40:29.9 & PL/EX & 0--350$^{a}$ & $135_{-8}^{+6}$ & $1.2 \pm 0.2$
 & $1.8 \times 10^{-7}$ & 20.0 / 24 & $141 \pm 6$ & $(3.4 \pm 0.3)
 \times 10^{-9}$ & 118.2 / 27\\
       &                         & PLE & 0--350 & {\boldmath $127 \pm
 5$} & {\boldmath $2.4 \pm 0.1$} & {\boldmath $1.3 \times 10^{-5}$$^{\dagger}$} &
 {\bf --} & {\boldmath $141 \pm 5$} & {\boldmath $2.8 \times 10^{-9}$$^{\dagger}$} & {\bf 22.1 / 26}\\\hline
060211A & 2006-02-11 & XRT & 232--312 & 0$^{\dagger}$ & $2.1 \pm 0.9$ 
 & $1.6 \times 10^{-5}$ & 5.0 / 6 & $127_{-39}^{+96}$ &
 $8.3_{-5.0}^{+12.2} \times 10^{-10}$ & 5.2 / 6\\
        & 09:39:59.9  & PL/EX & 0--913 & 72 & $2.3$ &
 $1.7 \times 10^{-5}$ & 43.8 / 29 & $81 \pm 2$ & 
 $(3.6 \pm 0.2) \times 10^{-9}$ &  333.0 / 32\\
        &             & PLE & 0--913 & {\bf 72} & {\boldmath $2.29 \pm 0.02$} & 
{\boldmath $1.4 \times 10^{-5}$$^{\dagger}$} & {\bf --} 
& {\boldmath $54 \pm 4$} & {\boldmath $3.1 \times 10^{-9}$$^{\dagger}$}
 & {\bf 39.9 / 31}\\\hline
060306  & 2006-03-06 & XRT & 97--147 & 0$^{\dagger}$ & $3.7 \pm 0.8$ &
 $2.2 \times 10^{-3}$ & 3.1 / 2 & $31_{-6}^{+9}$ &
 $1.7_{-1.0}^{+2.2} \times 10^{-9}$ & 3.3 / 2 \\
        & 00:49:09.3 & PL/EX & 35--256 & 40 & $2.3$ & $1.1 \times 10^{-6}$ & 31.4 / 4 & 
$17.8 \pm 0.7$ & $(3.1 \pm 0.7) \times 10^{-8}$ & 145.6 / 6\\
        &            & PLE & 35--256 & {\bf 40} & {\boldmath $2.33
 \pm 0.02$} & {\boldmath $1.0 \times 10^{-6}$$^{\dagger}$} & {\bf --} & 
{\boldmath $13 \pm 1$} & {\boldmath $2.5 \times 10^{-8}$$^{\dagger}$} &
 {\bf 25.4 / 5}\\\hline
060418 & 2006-04-18 & XRT & 178--400 & 0$^{\dagger}$ & $2.6 \pm 0.1$ &
 $3.1_{-1.7}^{+3.8} \times 10^{-4}$ & 27.1 / 28 & $101 \pm 6$ &
 $(2.0 \pm 0.3) \times 10^{-9}$ & 31.6 / 28\\
       & 03:05:49.2$^{b}$ & PL/EX & 108--797$^{c}$ & $146 \pm 2$ & 
$1.07 \pm 0.08$ & $2.2_{-0.8}^{+1.2} \times 10^{-8}$ & 25.0 / 30 & $262 \pm 23$ & 
$(3.0 \pm 0.5) \times 10^{-10}$ & 123.0 / 31\\
       &           & PLE & 108--797$^{b}$ & {\boldmath $148.7 \pm 0.3$} & 
{\boldmath $0.89 \pm 0.01$} & {\boldmath $7.5 \times 10^{-9}$$^{\dagger}$} & {\bf --} 
& {\boldmath $43 \pm 3$} & {\boldmath $1.5 \times 10^{-8}$$^{\dagger}$} & {\bf 23.4 / 30}\\\hline
060427 & 2006-04-27 & XRT & 148--198 & 0$^{\dagger}$ &
 $4.4 \pm 1.3$ & $4.8 \times 10^{-1}$ & 3.7 / 3 & $41_{-10}^{+20}$ & $5.1_{-0.3}^{+14.8}
 \times 10^{-9}$ & 4.0 / 3 \\
       &  11:43:01.0 & PL/EX  & 0--218 & 66 & $2.8$ & $3.7 \times
 10^{-5}$ & 3.5 / 3 & {\boldmath $47 \pm 2$} & {\boldmath $(2.9 \pm 0.4)
 \times 10^{-9}$} & {\bf 5.1 / 6}\\
       &             & PLE & 0--218 & 114 & $6.9$ & 2.7$^{\dagger}$ 
& -- & 47 & $2.8 \times 10^{-9}$$^{\dagger}$ & 3.3 / 5\\\hline
060428B & 2006-04-28 & XRT & 235--340 & 0$^{\dagger}$ & $5.0 \pm 0.7$ & 
 74 & 7.9 / 7 & $55_{-7}^{+10}$ & $4.1_{-2.1}^{+4.6} \times 10^{-9}$ & 9.4 / 7 \\
        &  08:54:15.2 & PL/EX & 0--340 & 229 & $0.7$ & $3.5 \times
 10^{-10}$ & 2.9 / 6 & {\boldmath $48 \pm 0.9$} & {\boldmath $8.9 \times
 10^{-9}$} & {\bf 10.4 / 8} \\
        &             & PLE & 0--340 & $236_{-2}^{+1}$ & $0.40 \pm
 0.05$ & $6.8 \times 10^{-11}$$^{\dagger}$ & 2.7 / 7 & $39 \pm 2$ & $1.0 \times
 10^{-8}$$^{\dagger}$ & 2.7 / 7\\\hline
060923C & 2006-09-23 & XRT & 205--529 & 0$^{\dagger}$ & $2.7 \pm 0.3$ &
 $1.7 \times 10^{-4}$ & 3.5 / 5 & $115_{-14}^{+17}$ & $4.0 \times
 10^{-10}$ & 5.4 / 5\\
        & 13:33:10.8 & PL/EX & 0--529 & {\boldmath $-3 \pm 2$} &
 {\boldmath $2.76 \pm 0.02$} & {\boldmath $2.0 \times
 10^{-4}$$^{\dagger}$} & {\bf 3.5 / 6} & $43 \pm 1$ & $(1.3 \pm 0.2)
 \times 10^{-8}$ & 45.3 / 6 \\
        &  & PLE & 0--529 & $-2.9 \pm 6$ & $2.42 \pm 0.03$ & $2.6
 \times 10^{-5}$$^{\dagger}$ & 3.2 / 5 & $33_{-10}^{+4}$ & $9.5 \times 10^{-9}$$^{\dagger}$ &
 3.2 / 5\\\hline
\enddata
\tablenotetext{\ddagger}{XRT: fitting only the XRT data with a PLO model and an EXP model, PL/EX: joint BAT and XRT fit with a PLO model and an EXP model, PLE: joint BAT and XRT fit with a PLEXP model.}
\tablenotetext{\dagger}{Fixed value.}
\tablenotetext{a}{All the BAT data points and the XRT data from $t_{0}$+224 s to 
$t_{0}$+350 s are used in the fit.}
 \tablenotetext{b}{Although {\it battblocks} found a time interval 
 which is 63 second before 
 $t_{0}$, we concluded that this interval is due to the contamination of
 Sco X-1 in the BAT field of view based on the BAT image analysis.}
\tablenotetext{c}{The fit to the XRT data is from $t_{0}$+108 s to $t_{0}$+148 s, 
and from $t_{0}$+797 s to $t_{0}$+400 s.  The BAT data points from $t_{0}$+146 s 
to $t_{0}$+156 s are also included in the fit.}

\end{deluxetable}

\newpage
\begin{deluxetable}{ccccc}
\tablecaption{XRT spectral parameters based on a joint fit to WT and PC 
 data using data above 2 keV. The error in the photon index is quoted at 
the 90\% confidence level.  \label{tab:xrt_spec_para}}
\tablewidth{0pt}
\tablehead{
\colhead{GRB} & 
\colhead{WT Fitting Range} & 
\colhead{PC Fitting Range} & 
\colhead{$\Gamma_{XRT}$} & 
\colhead{$\chi^{2}$/dof}\\
\colhead{} &
\colhead{[s]} &
\colhead{[s]} &
\colhead{} &
\colhead{}}
\startdata
050803  & 100.1--184.1   & 185.5--12976 & $1.9 \pm 0.2$ & 33.0 / 37\\
050814  & 166.6--384.9 & 386.5--14133 & $2.1 \pm 0.2$ & 33.5 / 35\\
050915B & 158.4--313.3 & 229.9--13791 & $1.8 \pm 0.3$ & 9.3 / 11\\
051109A & 131.4--214.3 & 3444--17540 & $2.2 \pm 0.2$ & 19.0 / 31\\
060109  & 109.8--200.2 & 201.6--12559 & $2.0 \pm 0.2$ & 23.7 / 21\\
060111B & 89.2--426.7 & 155.4--12655 & $2.2 \pm 0.3$ & 10.0 / 14 \\
060202  & 175.1--1025 & 1028--13027 & $1.99_{-0.04}^{+0.05}$ & 364.6 /  337\\
060211A & 137.1--330.9 & 332.5--13054 & $1.8 \pm 0.1$ & 68.6 / 66\\
060306  & 96.9--174.5 & 175.8--12874 & $2.40 \pm 0.03$ & 9.8 / 15\\
060418  & 166.1--697.7 & 699.9--12561 & $2.00_{-0.06}^{+0.07}$ & 211.7 / 209\\
060427  & 148.2--236.5 & 237.9--12691. & $2.0 \pm 0.4$  & 7.7 / 10\\
060428B & 235.6--440.8 & 442.7--12773 & $2.6 \pm 0.3$ & 16.6 / 17 \\
060923C & 204.6--267.4 & 268.7--12796 & $2.1 \pm 0.4$ & 3.0 / 7
\enddata
\end{deluxetable}

\newpage
\begin{deluxetable}{cccccccccccc|cc}
\tabletypesize{\scriptsize}
\rotate
\tablecaption{BAT prompt emission properties, and the estimated Bulk Lorentz 
factor, $\gamma_{0}$, and the radius of the external shock, R$_{0}$ based on 
the external shock model.  The XRT decay index based on the definition
 of \citet{takami2006}, $\alpha_{tail}$, and $\chi^{2}$ of the fit are also shown.  
Errors quoted the 68\% confidence level.  \label{tab:bat_grb_para}}
\tablewidth{0pt}
\tablehead{
\colhead{GRB} & 
\colhead{T$_{90}$} &
\colhead{Model$\star$} & 
\colhead{$\Gamma_{BAT}$} & 
\colhead{$\ep$} &
\colhead{S$_{E}$$^{a}$} &
\colhead{F$^{peak}_{E}$$^{b}$} &
\colhead{S3/S2$^{c}$} &
\colhead{z} &
\colhead{$T_{1/2}$} &
\colhead{$\gamma_{0}^{\ddagger}$} & 
\colhead{$R_{0}$} & 
\colhead{$\alpha_{tail}$} &
\colhead{$\chi^2$/d.o.f.}\\
\colhead{} &
\colhead{[s]} &
\colhead{} &
\colhead{} &
\colhead{[keV]} &
\colhead{} &
\colhead{} &
\colhead{} &
\colhead{} &
\colhead{} &
\colhead{} &
\colhead{[cm]} &
\colhead{} &
\colhead{}}
\startdata
050803 & 87.9 & PL & $1.39 \pm 0.07$ & -- & $22 \pm
 1$ & $8.1 \pm 0.7$ & $1.5 \pm 0.1$ & -- & 18 & 270 & $2.8 \times 10^{16}$ & -- & --\\
050814 & 140.6 & PL & $1.8 \pm 0.1$ & -- & $19 \pm 
 1$ & $6.2 \pm 1.2$ & $1.2 \pm 0.1$ & 5.3$^{1}$ & 45 & 240 & $3.0 \times 10^{16}$ & -- & --\\
050915B & 40.9 & CPL & $1.4 \pm 0.2$ & $60_{-5}^{+7}$ & $34 \pm 1$
 & $17 \pm 1$ & $1.05 \pm 0.04$ & -- & 24 & 242 & $3.0 \times 10^{16}$ & -- & --\\
051109A & 37.2 & PL & $1.5 \pm 0.1$ & -- & $22 \pm
 2$ & $29 \pm 3$ & $1.4 \pm 0.2$ & 2.346$^{2}$ & 15 & 287 & $2.7 \times 10^{16}$ & $0.59 \pm 0.05$ & 9.7 / 14\\
060109 & 115.4 & PL & $1.9 \pm 0.1$ & -- & $6.6 \pm
 0.6$ & $3.4 \pm 0.1$ & $1.0 \pm 0.2$ & -- & 23 & 246 & $3.0 \times 10^{16}$ & -- & --\\
060111B & 58.8 & PL & $1.0 \pm 0.1$ & -- & $16 \pm 1$ & 
$14 \pm 2$ & $2.0 \pm 0.1$ & -- & 9 & 350 & $2.4 \times 10^{16}$ &  $0.9 \pm 0.1$ & 1.0 / 3\\
060202 & 198.9 & PL & $1.8 \pm 0.1$ & -- & $22 \pm
 1$ & $3.7 \pm 0.8$ & $1.2 \pm 0.1$ & -- & 98 & 143 & $4.3 \times 10^{16}$ & -- & --\\
060211A & 126.3 & CPL & $0.9 \pm 0.3$ & $58_{-5}^{+8}$ & $16 \pm
 1$ & $3.3 \pm 0.1$ & $1.1 \pm 0.1$ & -- & 24 & 242 & $3.0 \times 10^{16}$ 
& $1.3 \pm 0.2$ & 1.0 / 2\\
060306 & 61.2 & PL & $1.80 \pm 0.05$ & -- & $21 \pm
 1$ & $47 \pm 2$ & $1.1 \pm 0.1$ & -- & 46 & 190 & $3.5 \times 10^{16}$ & $1.2 \pm 0.2$ & 2.4 / 3\\
060418 & 103.1 & PL & $1.64 \pm 0.03$ & -- & $80
 \pm 1$ & $49 \pm 2$ & $1.28 \pm 0.03$ & 1.489$^{3}$ & 21 & 226 & $3.1 \times 10^{16}$  & -- & --\\
060427 & 64 & PL & $1.9 \pm 0.2$ & -- & $5.0 \pm 0.5$ & 
$1.7 \pm 0.7$ & $1.1 \pm 0.2$ & -- & 32 & 217 & $3.2 \times 10^{16}$ & -- & --\\
060428B & 57.9 & PL & $2.6 \pm 0.1$ & -- & $8.2 \pm 0.5$ & 
$3.4 \pm 0.6$ & $0.7 \pm 0.1$ & -- & 27 & 232 & $3.1 \times 10^{16}$  & $3.0 \pm 0.3$ & 4.7 / 8\\
060923C & 75.8 & PL & $2.3 \pm 0.1$ & -- & $16 \pm
 1$ & $5.0 \pm 1.5$ & $0.8 \pm 0.1$ & -- & -- & -- & -- & -- & --\\
\enddata
\tablenotetext{\star}{PL: power-law (dN/dE $\sim$ E$^{-{\Gamma_{BAT}}}$),
 CPL: cutoff power-law (dN/dE $\sim$ E$^{-{\Gamma_{BAT}}}$ $\exp(-(2-\Gamma_{BAT})E/\ep)$)} 
\tablenotetext{\ddagger}{Calculated bulk Lorentz factor assuming
 $E_{0}$/n = $1 \times 10^{52}$ erg cm$^{3}$.}
\tablenotetext{a}{Energy fluence in the 15--150 keV band [10$^{-7}$ erg cm$^{-2}$]}
\tablenotetext{b}{1-s peak energy flux in the 15--150 keV band [10$^{-8}$ erg cm$^{-2}$ s$^{-1}$]}
\tablenotetext{c}{Fluence ratio between S3(50--150 keV)/S2(25--50 keV)}
\tablenotetext{1}{\citet{jakobsson2006}}
\tablenotetext{2}{\citet{quimby2005}}
\tablenotetext{3}{\citet{dupree2006}}
\end{deluxetable}

\newpage

\newpage
\begin{figure}
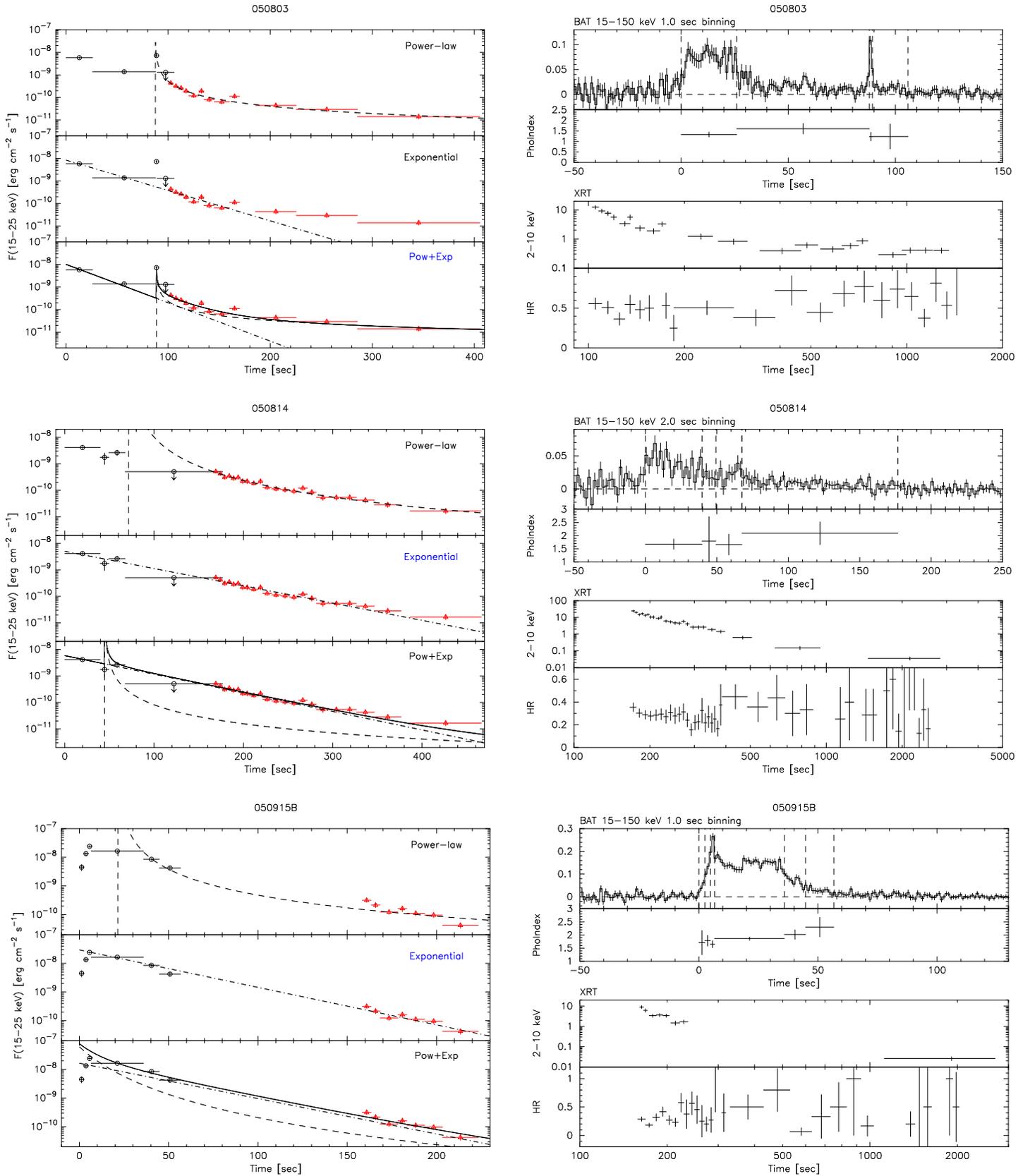

\centerline{
\includegraphics[width=7cm,angle=-90]{f2a.eps}
\hspace{0.5cm}
\includegraphics[width=7cm,angle=-90]{f2b.eps}}
\vspace{0.5cm}
\centerline{
\includegraphics[width=7cm,angle=-90]{f2c.eps}
\hspace{0.5cm}
\includegraphics[width=7cm,angle=-90]{f2d.eps}}
\vspace{0.5cm}
\centerline{
\includegraphics[width=7cm,angle=-90]{f2e.eps}
\hspace{0.5cm}
\includegraphics[width=7cm,angle=-90]{f2f.eps}}
\caption{The BAT and XRT composite light curve.  See text for details
 (section 3).}
\label{fig:bat_xrt_lc1}
\end{figure}

\newpage
\begin{figure}
\centerline{
\includegraphics[width=7cm,angle=-90]{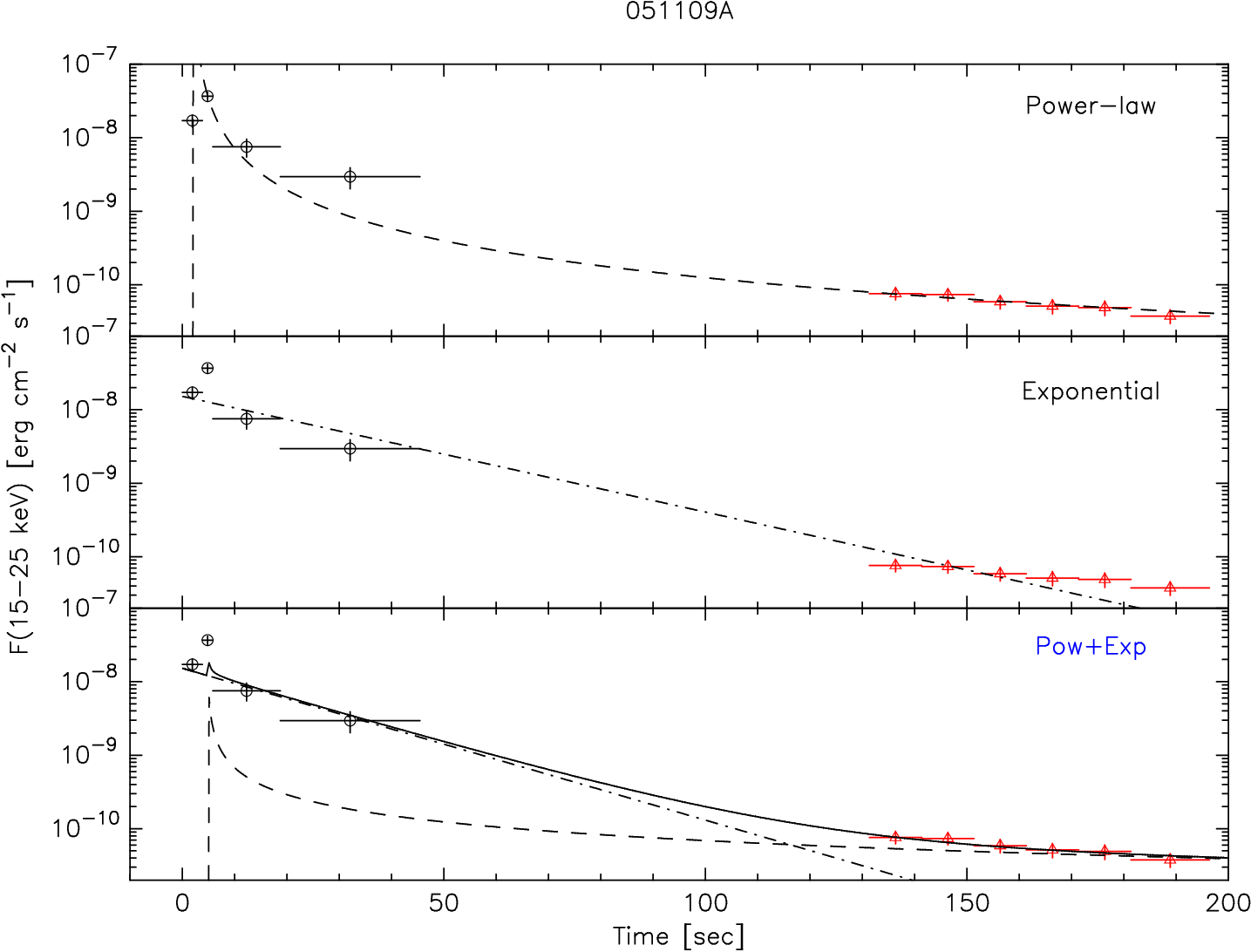}
\hspace{0.5cm}
\includegraphics[width=7cm,angle=-90]{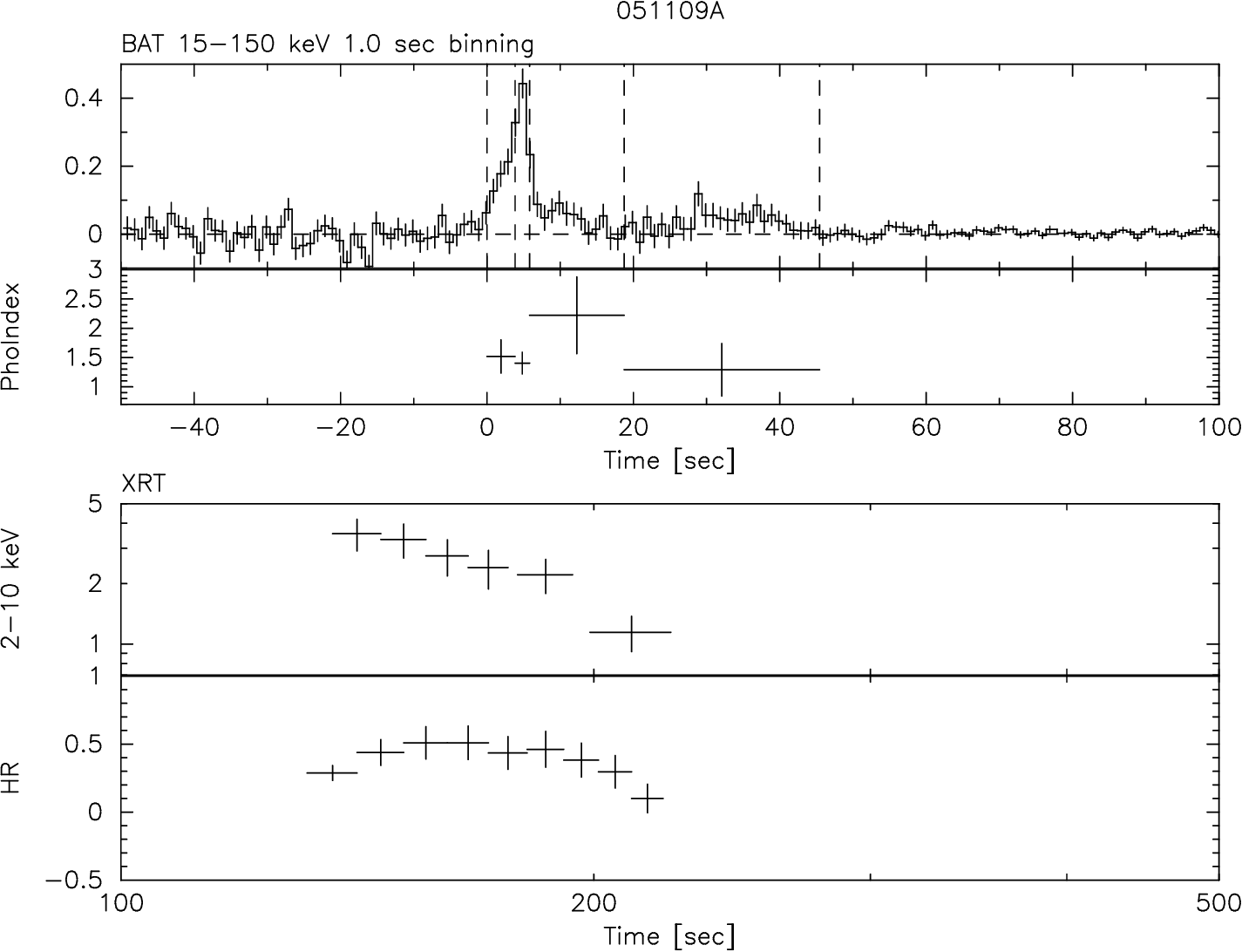}}
\vspace{0.5cm}
\centerline{
\includegraphics[width=7cm,angle=-90]{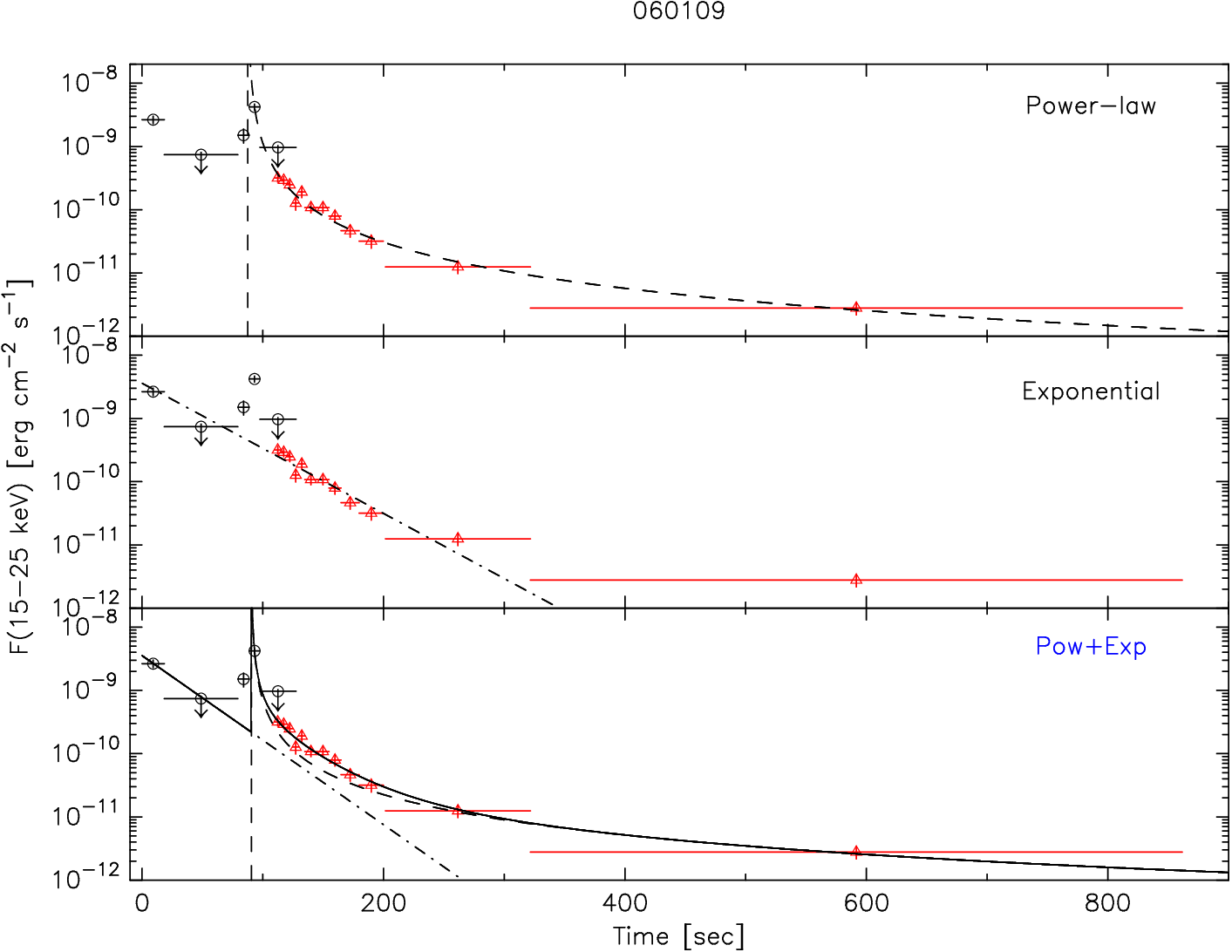}
\hspace{0.5cm}
\includegraphics[width=7cm,angle=-90]{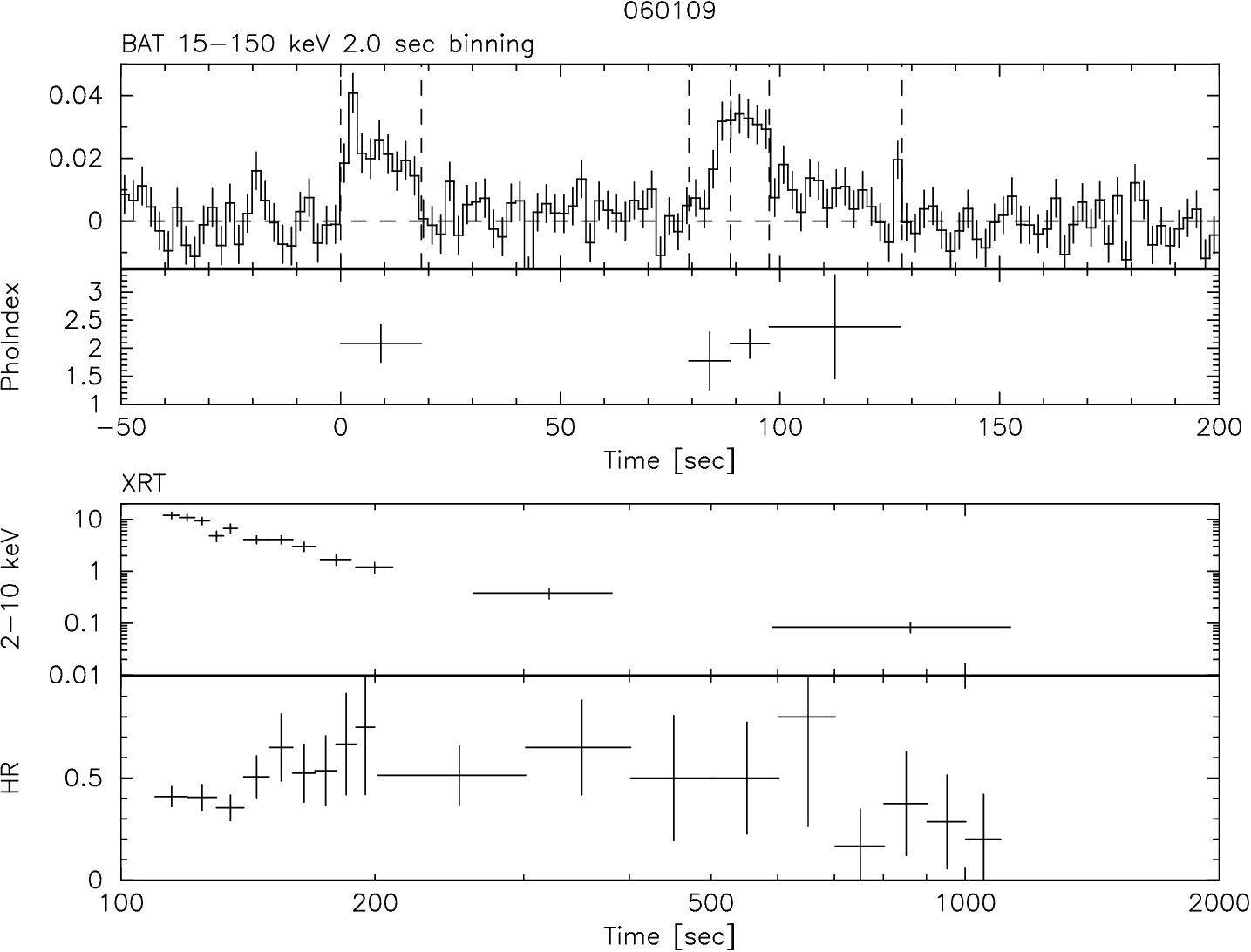}}
\vspace{0.5cm}
\centerline{
\includegraphics[width=7cm,angle=-90]{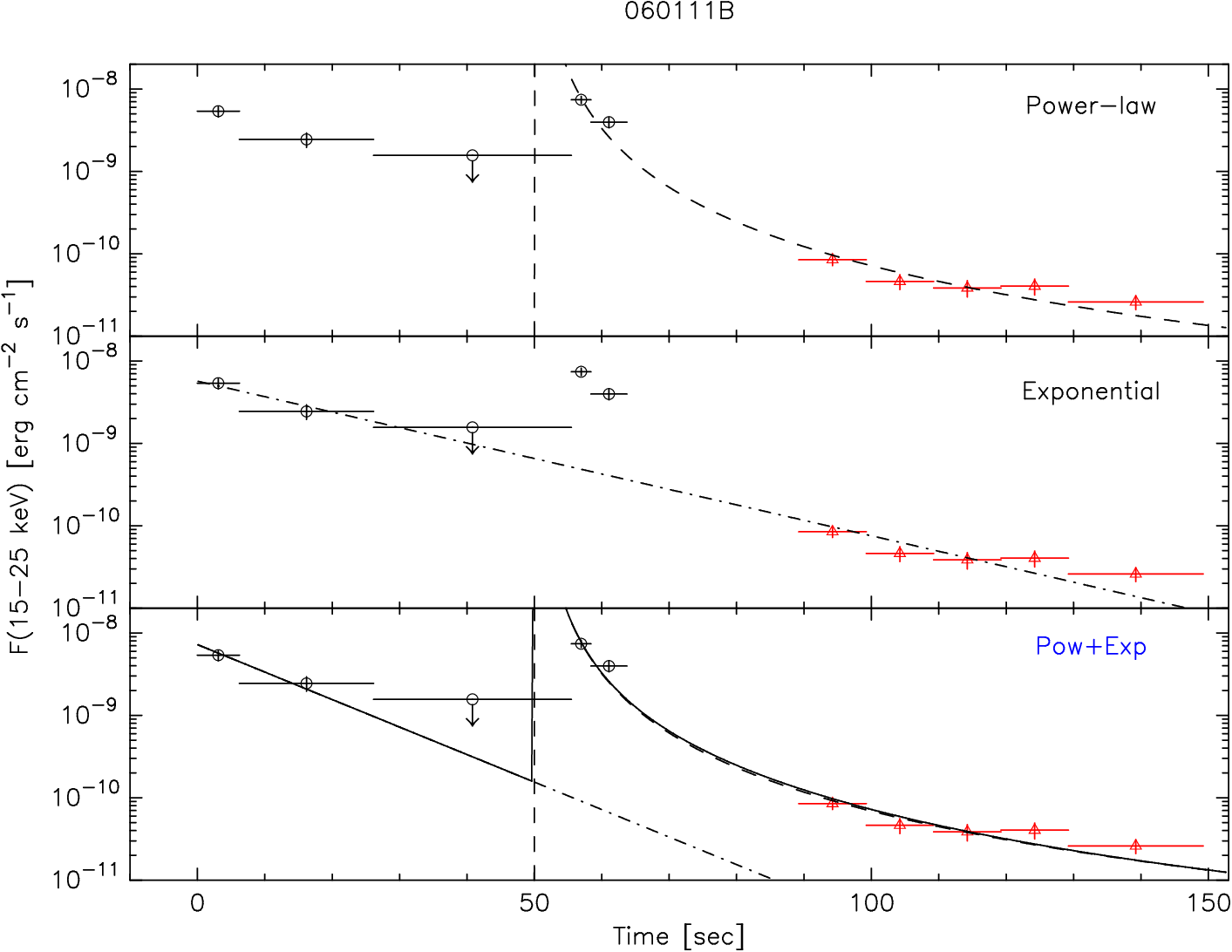}
\hspace{0.5cm}
\includegraphics[width=7cm,angle=-90]{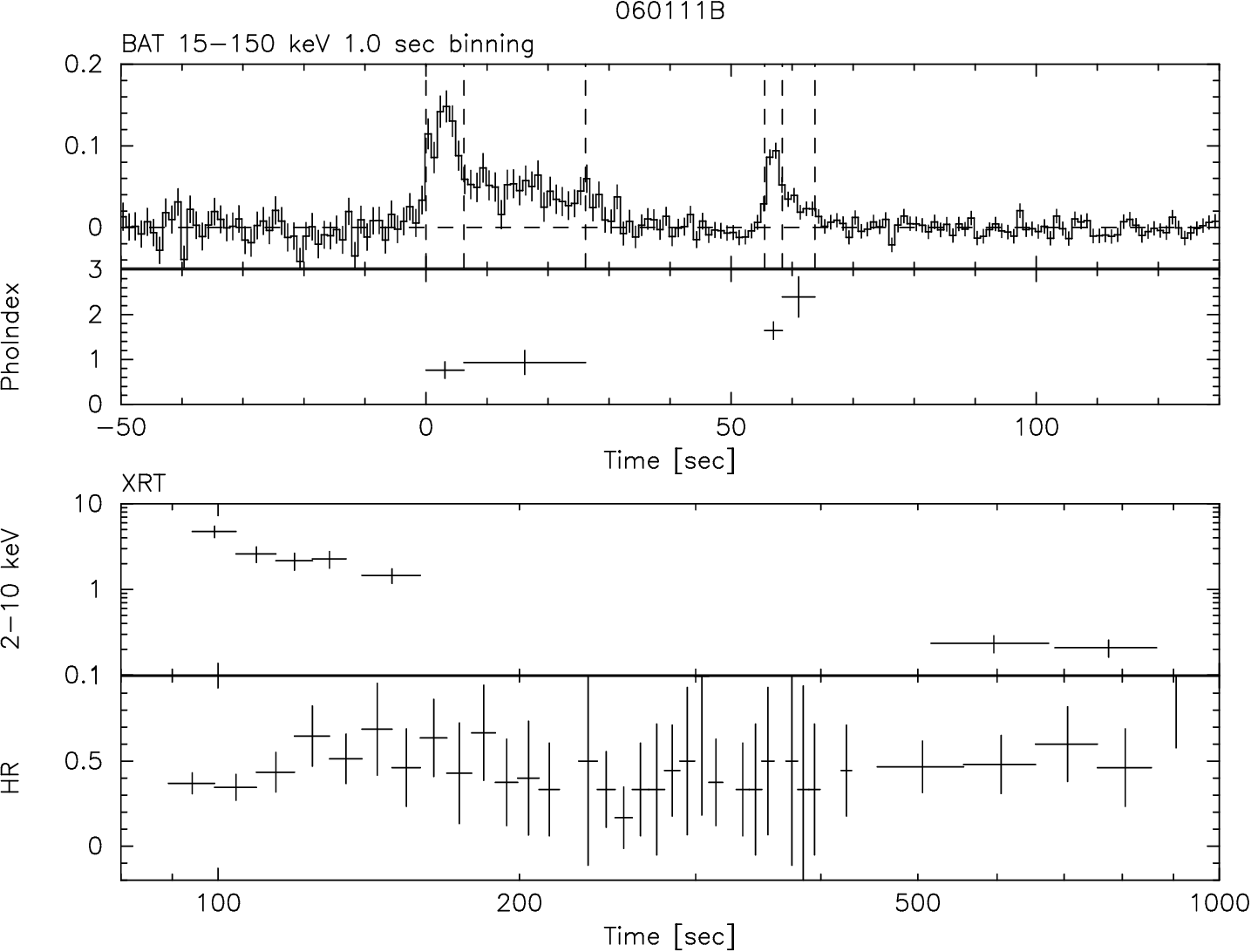}}
\caption{continued.}
\label{fig:bat_xrt_lc2}
\end{figure}

\newpage
\begin{figure}
\centerline{
\includegraphics[width=7cm,angle=-90]{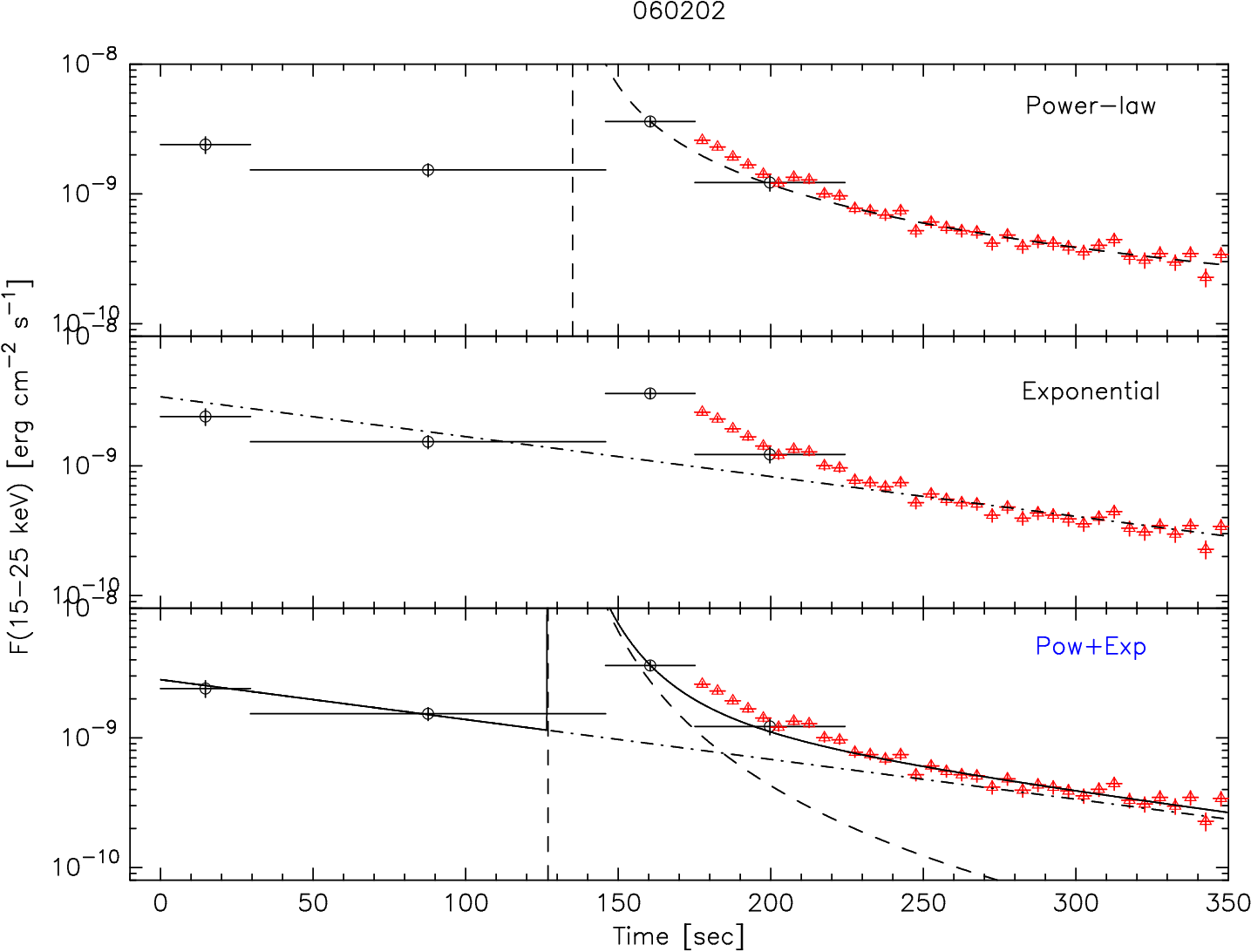}
\hspace{0.5cm}
\includegraphics[width=7cm,angle=-90]{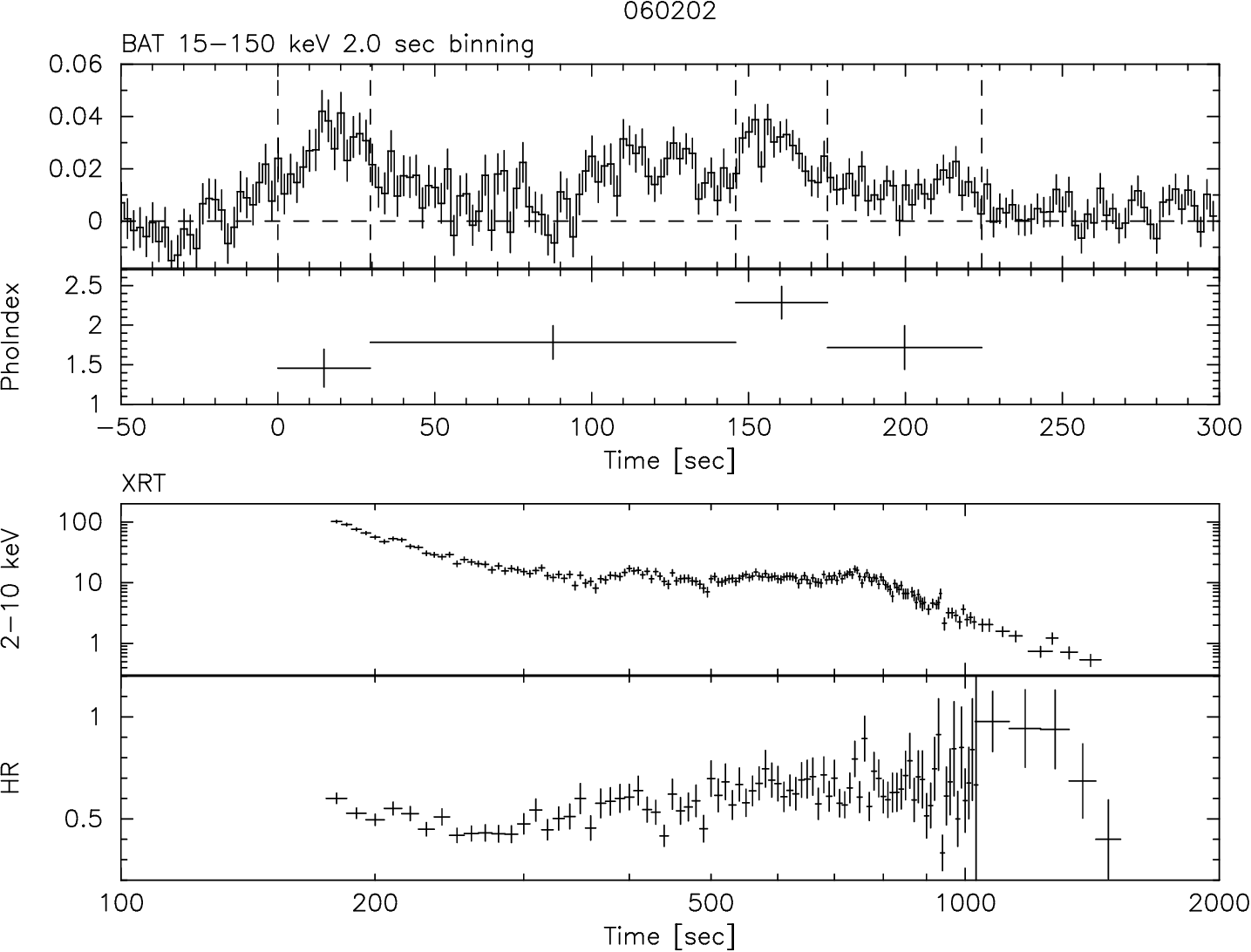}}
\vspace{0.5cm}
\centerline{
\includegraphics[width=7cm,angle=-90]{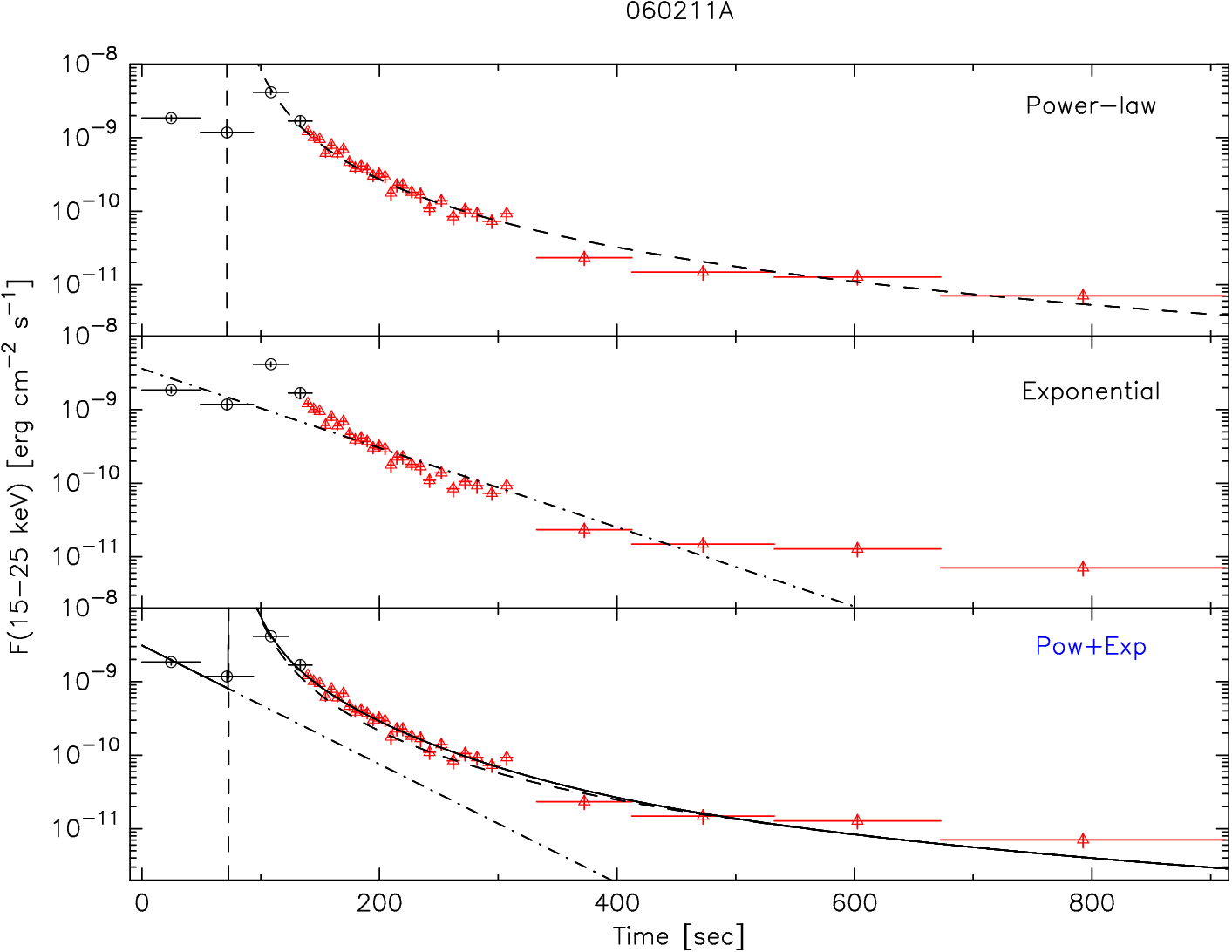}
\hspace{0.5cm}
\includegraphics[width=7cm,angle=-90]{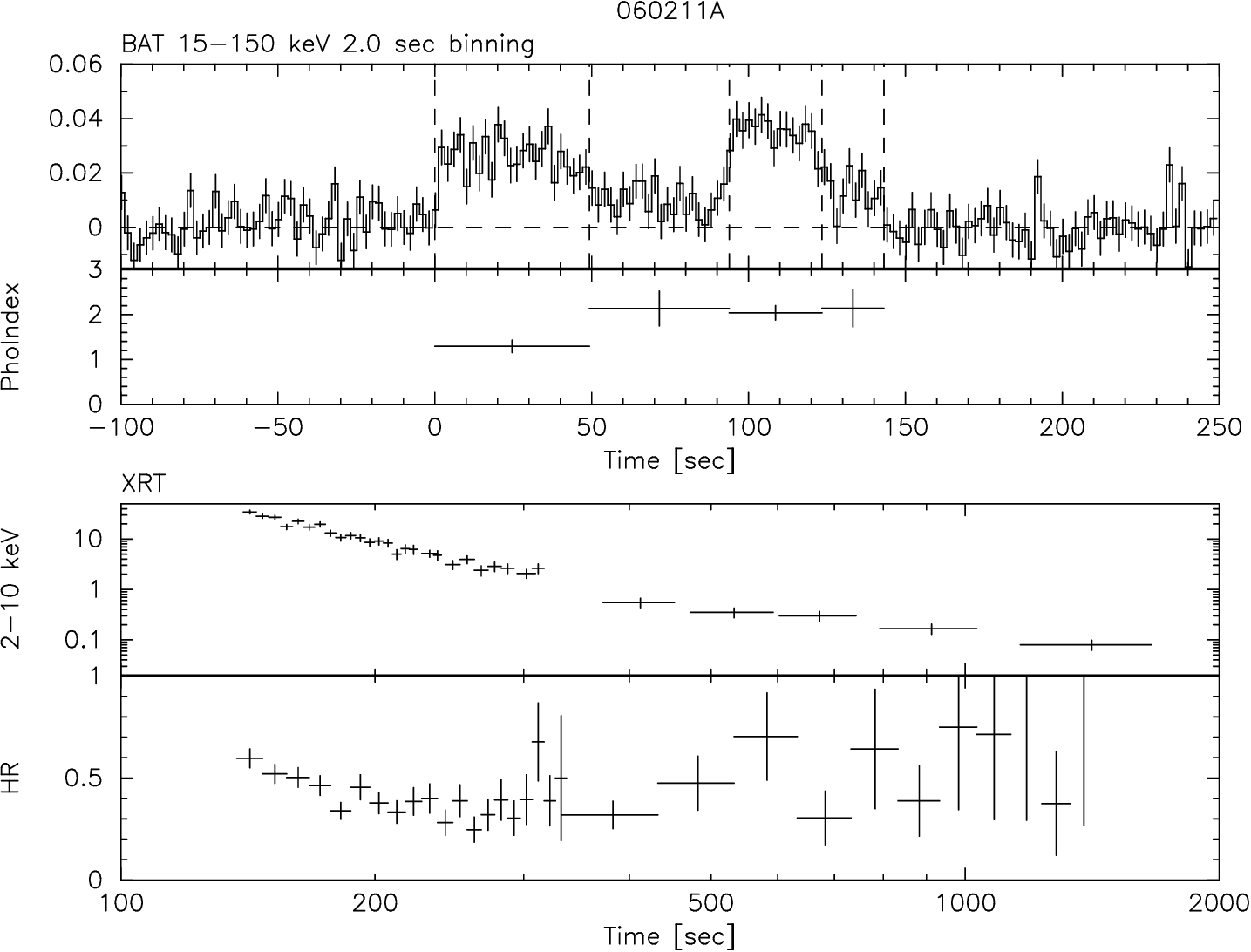}}
\vspace{0.5cm}
\centerline{
\includegraphics[width=7cm,angle=-90]{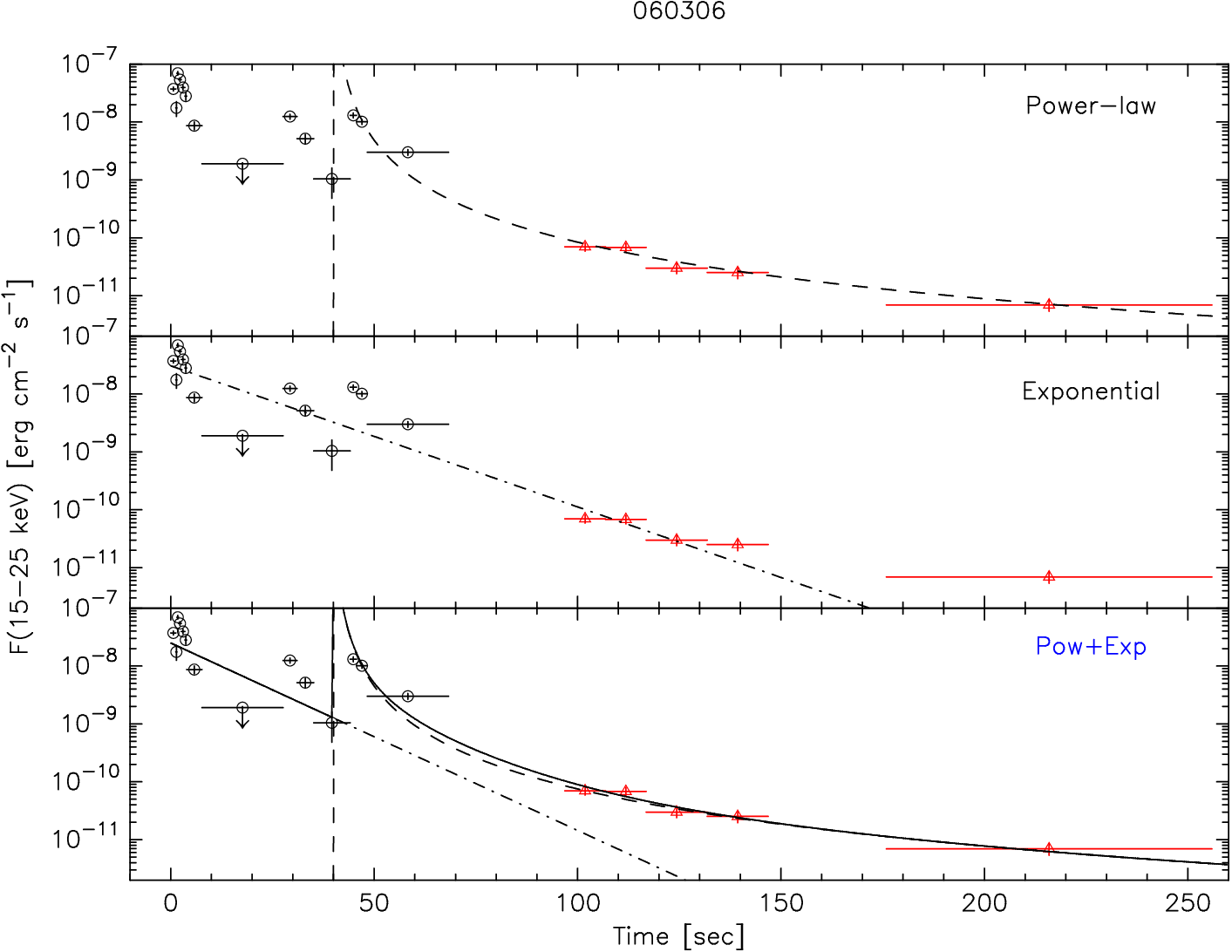}
\hspace{0.5cm}
\includegraphics[width=7cm,angle=-90]{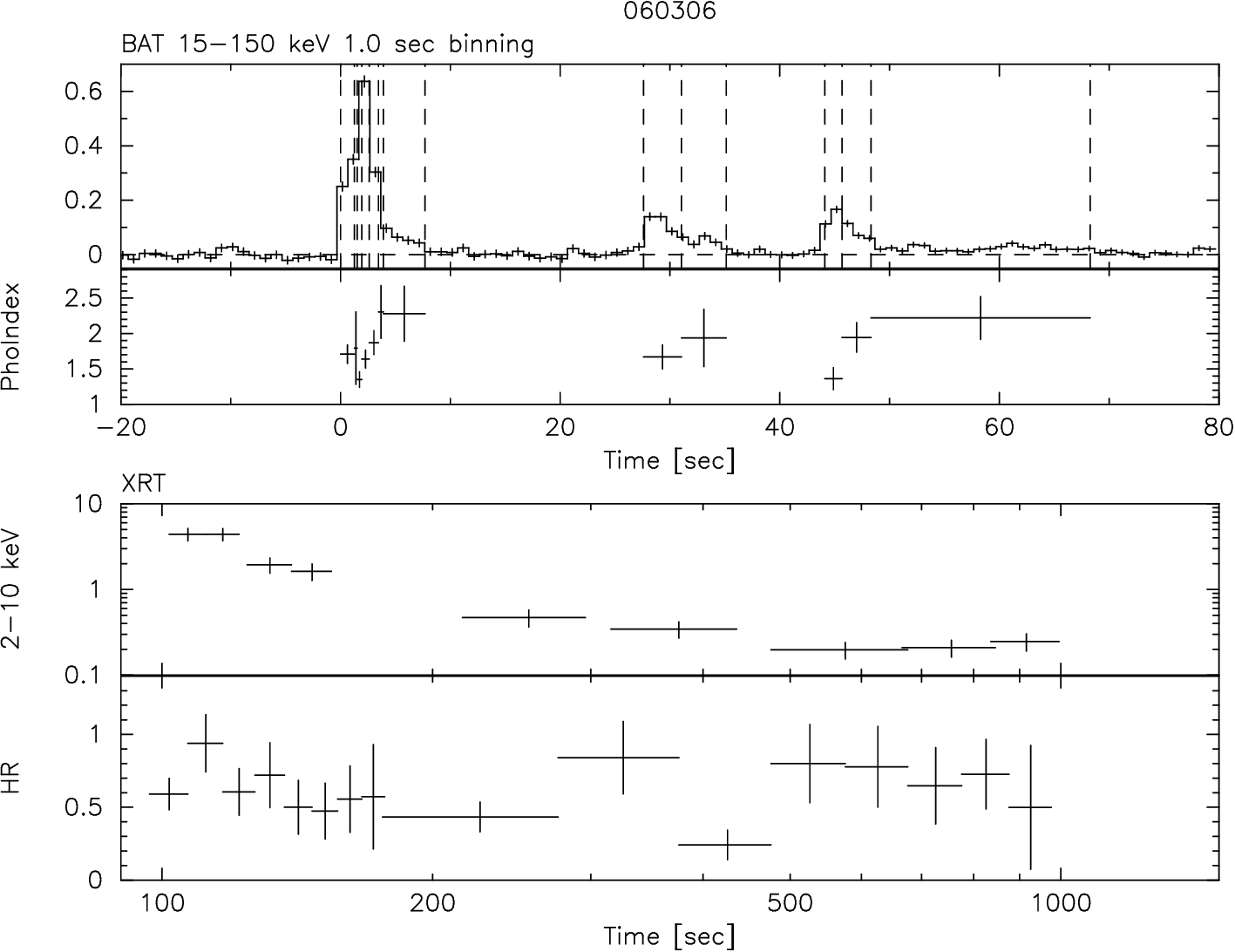}}
\vspace{0.5cm}
\caption{continued.}
\label{fig:bat_xrt_lc3}
\end{figure}

\newpage
\begin{figure}
\centerline{
\includegraphics[width=7cm,angle=-90]{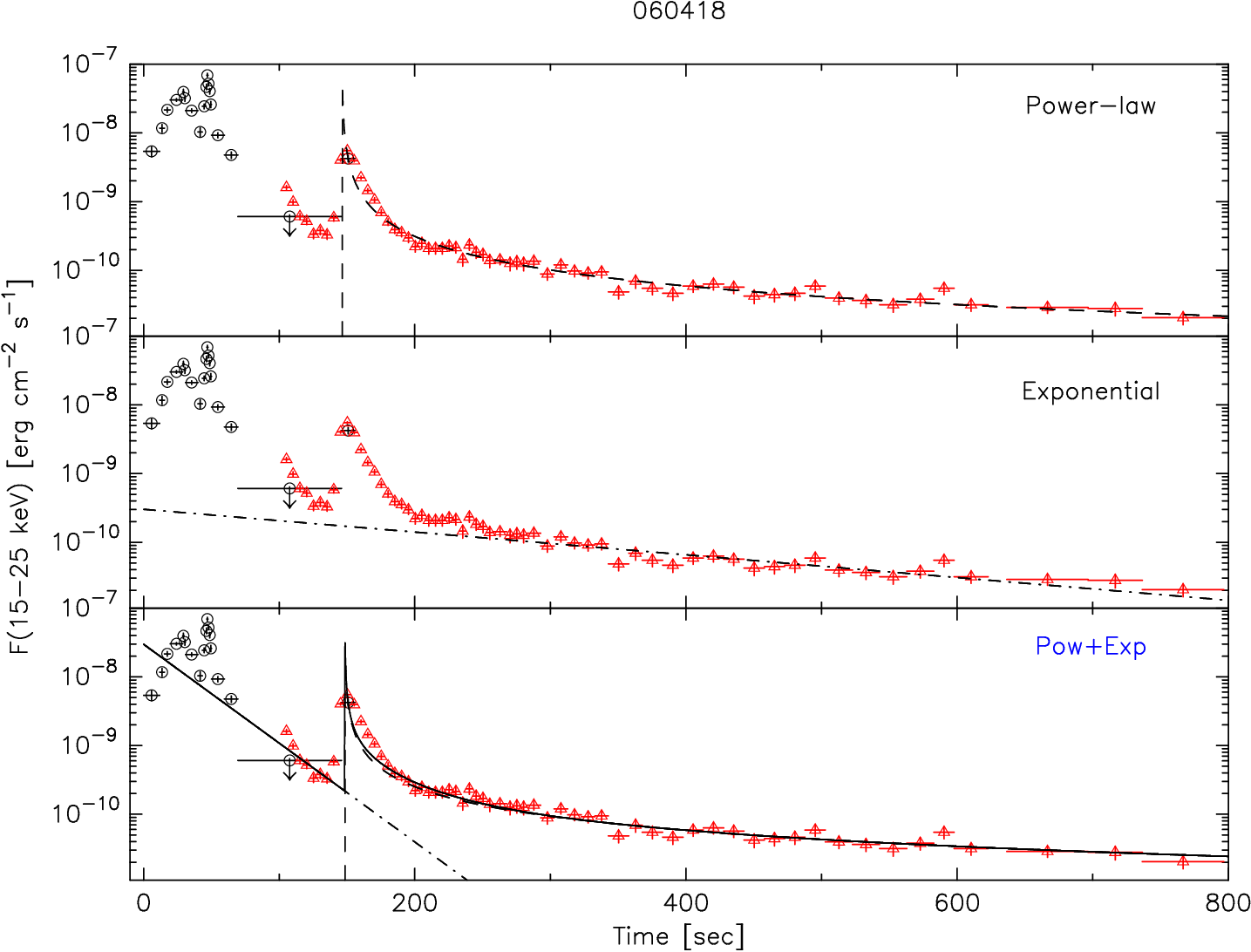}
\hspace{0.5cm}
\includegraphics[width=7cm,angle=-90]{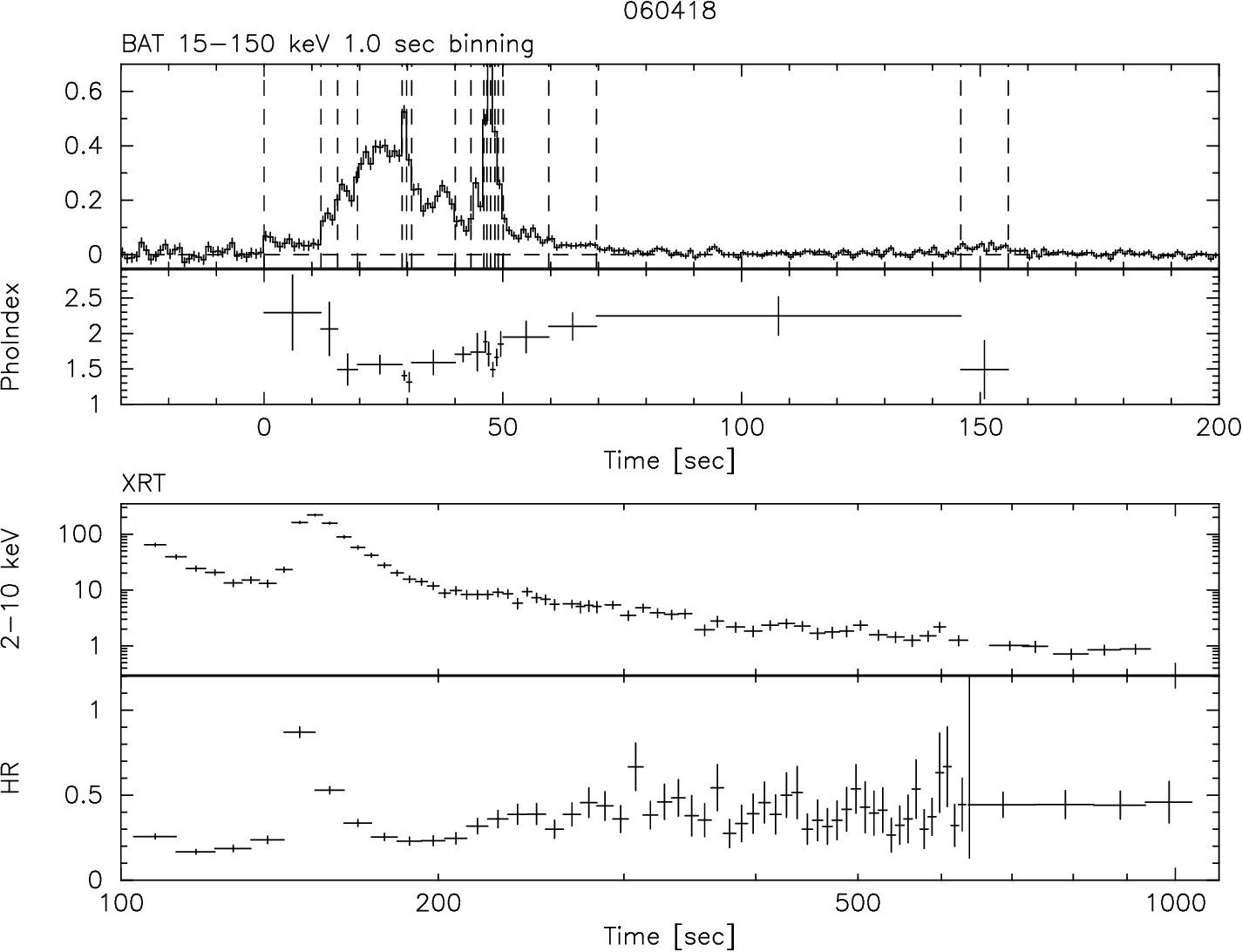}}
\vspace{0.5cm}
\centerline{
\includegraphics[width=7cm,angle=-90]{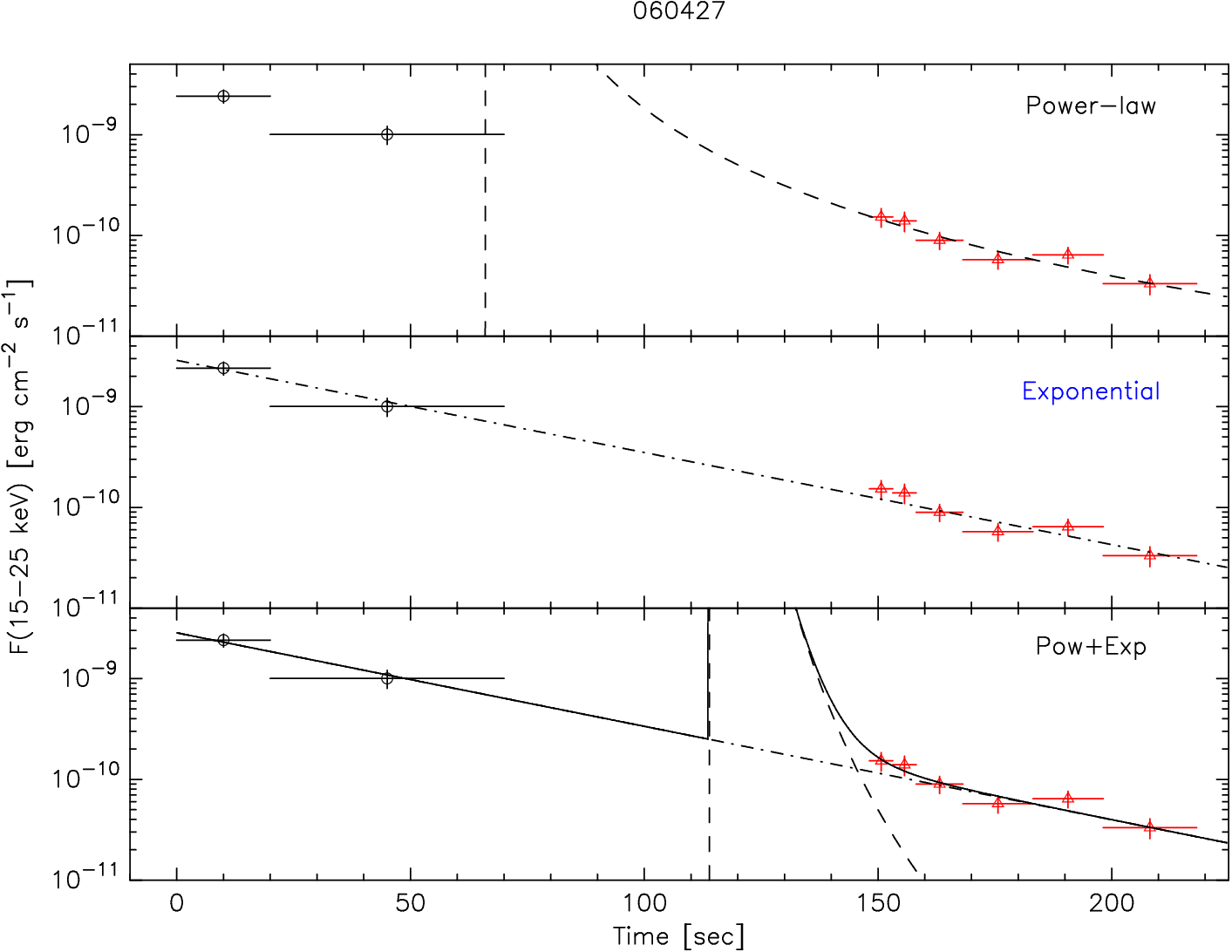}
\hspace{0.5cm}
\includegraphics[width=7cm,angle=-90]{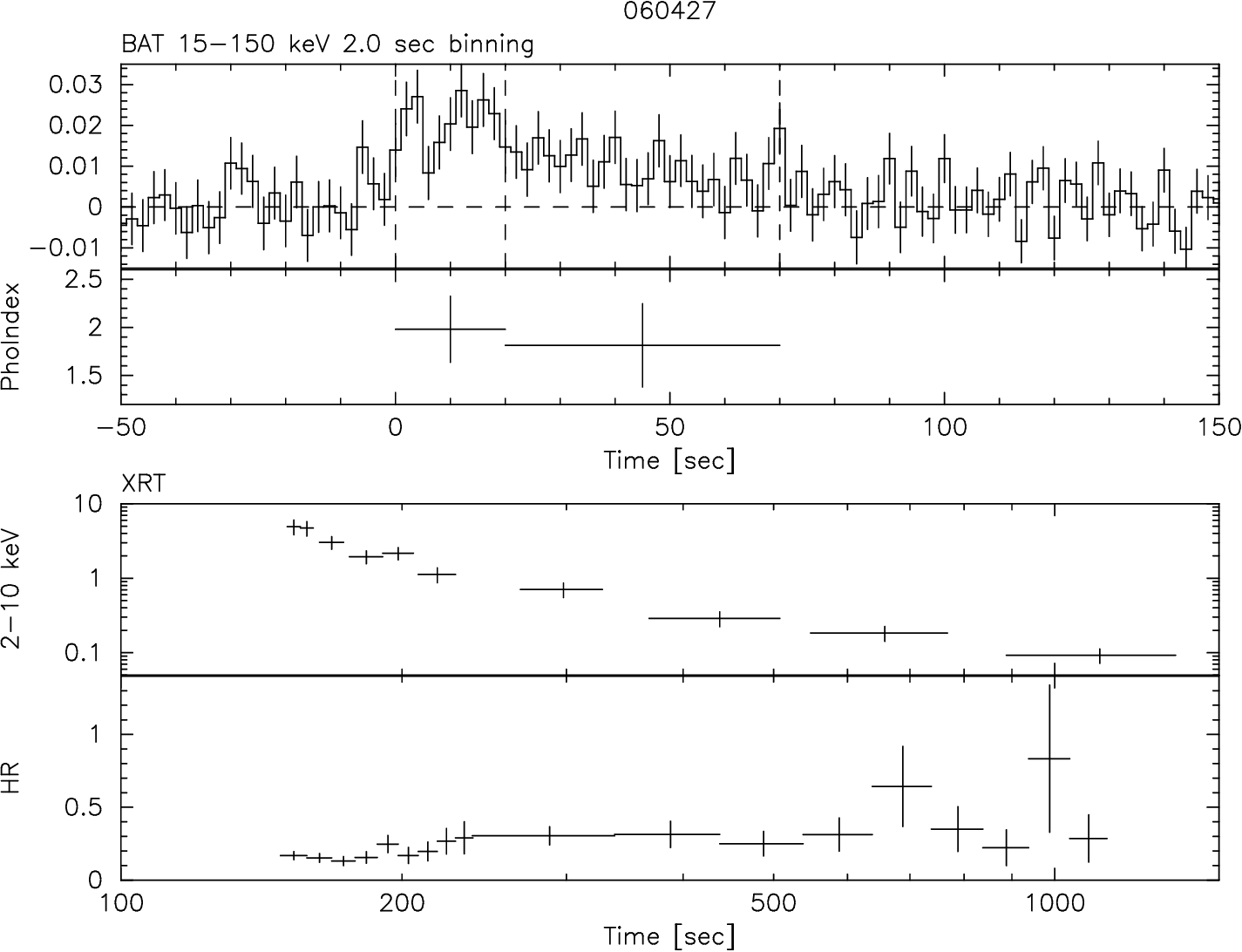}}
\vspace{0.5cm}
\centerline{
\includegraphics[width=7cm,angle=-90]{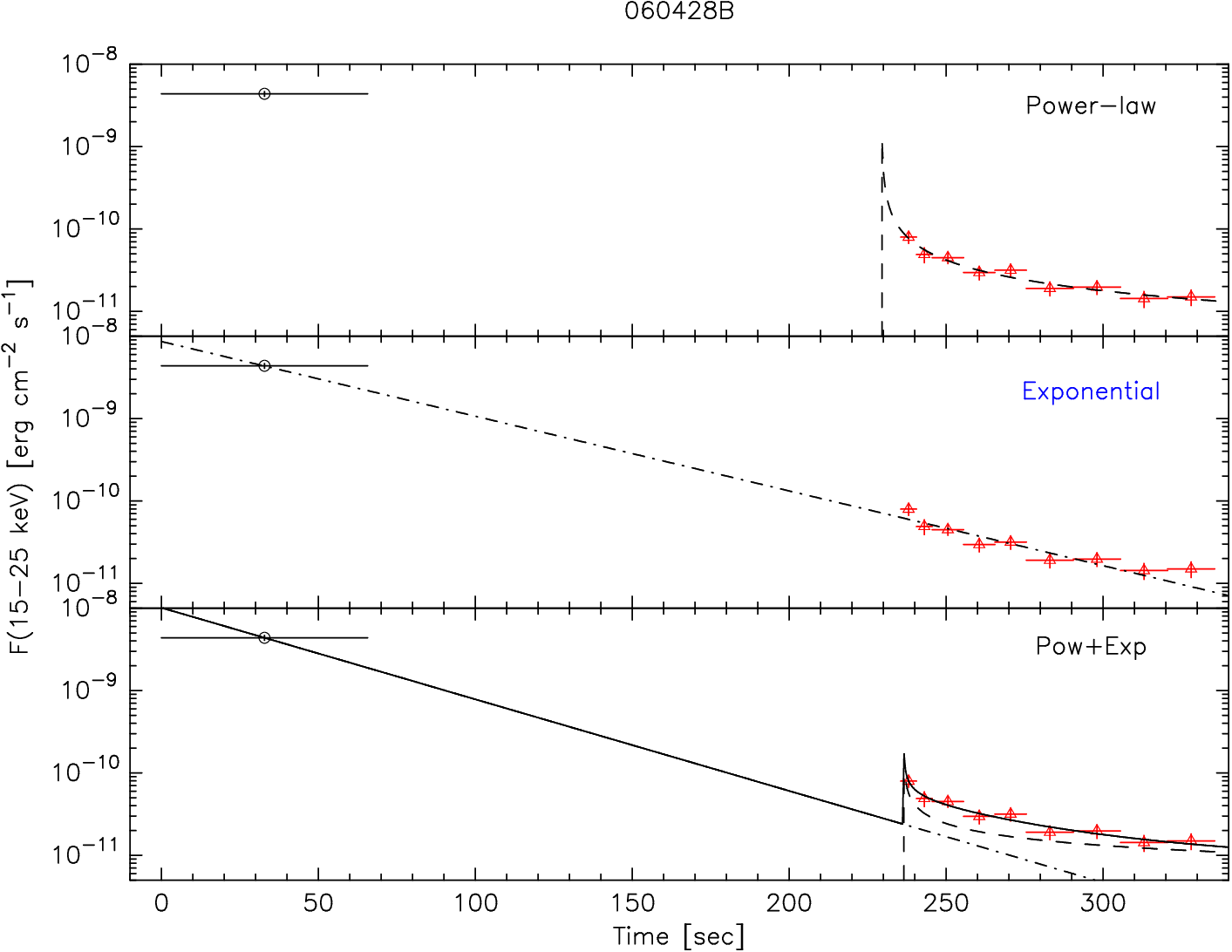}
\hspace{0.5cm}
\includegraphics[width=7cm,angle=-90]{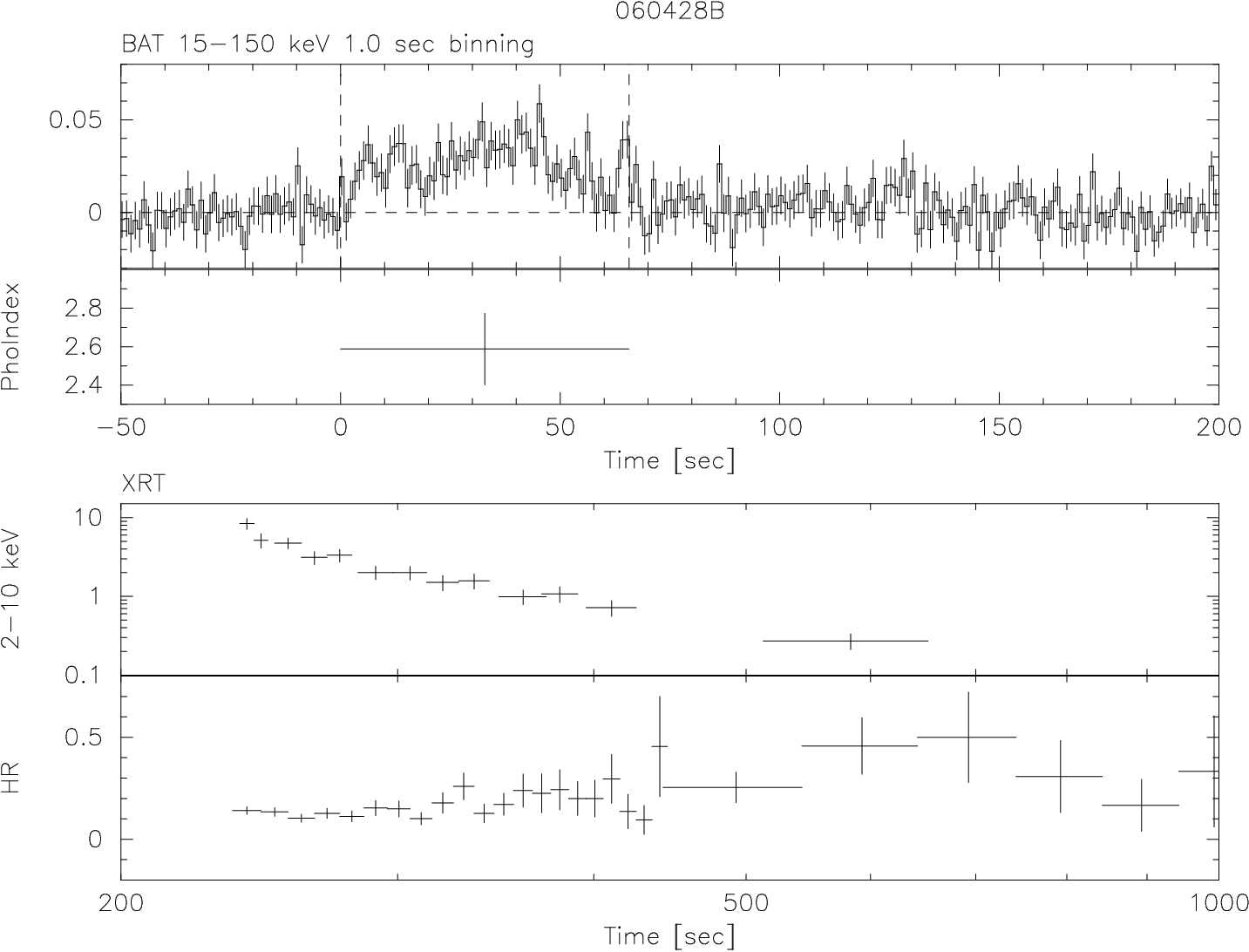}}
\caption{continued.}
\label{fig:bat_xrt_lc4}
\end{figure}

\newpage
\begin{figure}
\centerline{
\includegraphics[width=7cm,angle=-90]{f6a.eps}
\hspace{0.5cm}
\includegraphics[width=7cm,angle=-90]{f6b.eps}}
\caption{continued.}
\label{fig:bat_xrt_lc5}
\end{figure}

\newpage

\newpage

\newpage

\newpage






\clearpage

\end{document}